\documentclass[aip,reprint]{revtex4-1}
\usepackage{graphicx}
\usepackage{ulem}
\usepackage{color}

\begin{document}

\title{Knotting probability of self-avoiding polygons under a topological constraint}

\author{Erica Uehara and Tetsuo Deguchi}

%\begin{center}  
\affiliation{
Department of Physics, Faculty of Core Research, 
Ochanomizu University \\
2-1-1 Ohtsuka, Bunkyo-ku, Tokyo 112-8610, Japan} 
%\end{center} 

\begin{abstract}
We define the knotting probability of a knot $K$ by the probability for a random polygon (RP) or self-avoiding polygon (SAP) of $N$ segments having the knot type $K$. We show fundamental and generic properties of the knotting probability particularly its dependence on the excluded volume.  We investigate them for the SAP consisting of hard cylindrical segments of unit length and radius $r_{\rm ex}$. For various prime and composite knots we numerically show that a compact formula describes the knotting probabilities for the cylindrical SAP as a function of segment number $N$ and radius $r_{\rm ex}$.  It connects the small-$N$ to the large-$N$ behavior and even to lattice knots in the case of large values of radius.  As the excluded volume increases the maximum of the knotting probability decreases for prime  knots except for the trefoil knot. If it is large,   the trefoil knot and its descendants are dominant among the nontrivial knots in the SAP. From the factorization property of the knotting probability we derive  a relation among the estimates of a fitting parameter for all prime knots,  which suggests the local knot picture. Here we remark that the cylindrical SAP gives a model of  circular DNA which are negatively charged and semiflexible, where radius $r_{\rm ex}$ corresponds to the screening length.
\end{abstract}
\maketitle

\newpage
%***********************************************************************
% Section:
% Introduction
%***********************************************************************
\section{Introduction}

Statistical and dynamical properties of ring polymers under a topological constraint have attracted much interest in various branches of physics, chemistry and biology \cite{Kramers,Semlyen,Bates}. The topology of a ring polymer in solution is specified by a knot type (Fig. 1). Ring polymers with trivial topology are observed in nature such as circular DNA \cite{Vinograd}. Moreover, DNA with many knot types have been derived in experiments \cite{Nature-trefoil,DNAknots}. Topological structures related to knots or pseudo-knots have been discussed in association with protein folding \cite{Taylor}.  Naturally occurring proteins whose ends connected to give a circular topology has been recently discovered \cite{Craik}. Furthermore, a molecular knot with eight crossings in nanoscale has been successfully synthesized quite recently \cite{Woltering}.   Due to novel developments in experimental techniques during the last decade, ring polymers are now effectively synthesized in chemistry \cite{Tezuka2000,Tezuka2001,Grubbs,Takano05,Takano07,Grayson,Tezuka2010,Tezuka2011,Tezuka-book}.

We define the knotting probability of a knot by the probability for a  random polygon (RP) or self-avoiding polygon (SAP) consisting of $N$ segments having the given knot type.  It plays a fundamental role in the topological properties of ring polymers in solution. For instance, the mean-square radius of gyration of a knotted ring polymer depends not only on the knot type but also on the characteristic length of the knotting probability, which will be defined later in the paper.  The knotting probabilities have been studied for some models of RP and SAP through numerical simulations \cite{Vologodskii,Michels-Wiegel,Janse van Rensburg,Koniaris-Muthukumar,knotP,JKTR,TD95,Deguchi-Tsurusaki1997,Orlandini1996,Orlandini1998,PLA2000,Katritch00,Yao,Marcone,Stella,Rechnitzer,Tubiana,UD2015},   rigorous methods \cite{Sumners-Whittington,Pippenger} and DNA experiments \cite{Rybenkov,Shaw-Wang,Plesa}.

The knotting probabilities have been measured in experiments, first by performing the reaction process of randomly closing nicked circular DNA \cite{Rybenkov,Shaw-Wang}. In the researches circular DNA are modeled as SAP consisting of impenetrable cylinders, and the knotting probability is evaluated in the simulation for segment number $N$ up to 60. The results are compared with the experiments where segment number $N$ is rather small such as $N$ less than 30.   However, the knotting probability of large circular DNA such as 166 kbp has been measured recently in solid-state nanopore experiments \cite{Plesa}. Thus, the knotting probability for SAP with a large segment number $N$ such as $N=500$ can be systematically investigated in experiments.

In the paper we show fundamental and generic properties of the knotting probability of a given knot,  in particular, how it depends on the excluded volume. In order to investigate the excluded-volume effect on topological properties systematically,  we introduce an off-lattice model of  SAP which give random configurations of a cyclic sequence of  $N$ cylindrical segments  of radius $r_{\rm ex}$ where a given pair of segments do not overlap 
except for neighboring ones.  
We describe the knotting probability as a function of not only segment number $N$ 
but also radius $r_{\rm ex}$ by introducing a compact formula with four fitting parameters. 
The generic properties  presented in the paper should be useful for studying the knotting probability with experiments in various different fields.

We show numerically how the fitting parameters for 
expressing the knotting probability depend on cylindrical radius $r_{\rm ex}$, i.e., the excluded-volume parameter.    
We also show  that the four-parameter formula describes the knotting probability very well for various knots over a wide range of segment number $N$.   
The simulation results of the cylindrical SAP with several values of cylindrical radius lead to a systematic and unifying viewpoint on the knotting probability for many different models of ring polymers in solution. For instance, the knotting probability ratio is consistent with that of lattice SAP if cylindrical radius $r_{\rm ex}$ is large such as satisfying $2 r_{\rm ex}=1/4$: the diameter of cylindrical segments is given by one fourth of the bond length. Moreover, we show that if the cylindrical radius is large the trefoil knot and its descendants are dominant among the nontrivial knots appearing in an ensemble of SAP. We also show that the maximum of the knotting probability of the trefoil knot slightly increases with respect to the cylindrical radius, while those of other prime knots decrease exponentially with respect to it.  
Here we remark that the dependence of the knotting probability on the radius of cylindrical segments shown in the present study is consistent with the previous small-$N$ results  \cite{Rybenkov,Shaw-Wang} and generalizes them into those of the large $N$ case.

The cylindrical SAP model employed in the present research  generates  random sequences of impenetrable cylinders of unit length with radius $r_{\rm ex}$ where neighboring pairs of cylindrical segments can overlap while other pairs do not \cite{UD2015}. Here we recall that it is a SAP model of semi-flexible ring polymers such as circular DNA \cite{Rybenkov,Shaw-Wang}. In the model the radius $r_{\rm ex}$ corresponds to the screening length or the length scale of screening effect due to counter ions surrounding DNA  \cite{Schellman,Stigter,Rybenkov,Shaw-Wang}.  DNA are negatively charged polyelectrolytes and the screening effect of counter ions may be nontrivial \cite{Schellman,Stigter}. We assume that DNA chain is hard to bend due to electrostatic repulsive forces and hence DNA are approximated as a sequence of thin long cylinders where some fraction of counter ions are bound to DNA due to the Manning condensation \cite{LeBret}. The effective thickness of DNA molecules is determined by the concentration of counter ions in solution \cite{Stigter}.  Typically, the bare radius of DNA corresponds to the radius $r_{\rm ex}=0.01$ in the case of cylindrical segments of unit length \cite{Grosberg-book}.

%-----------------------------------------------------------------------
% Figure 1. 
\begin{figure}[htbp]
\begin{center}
  \includegraphics[width=0.8\hsize]{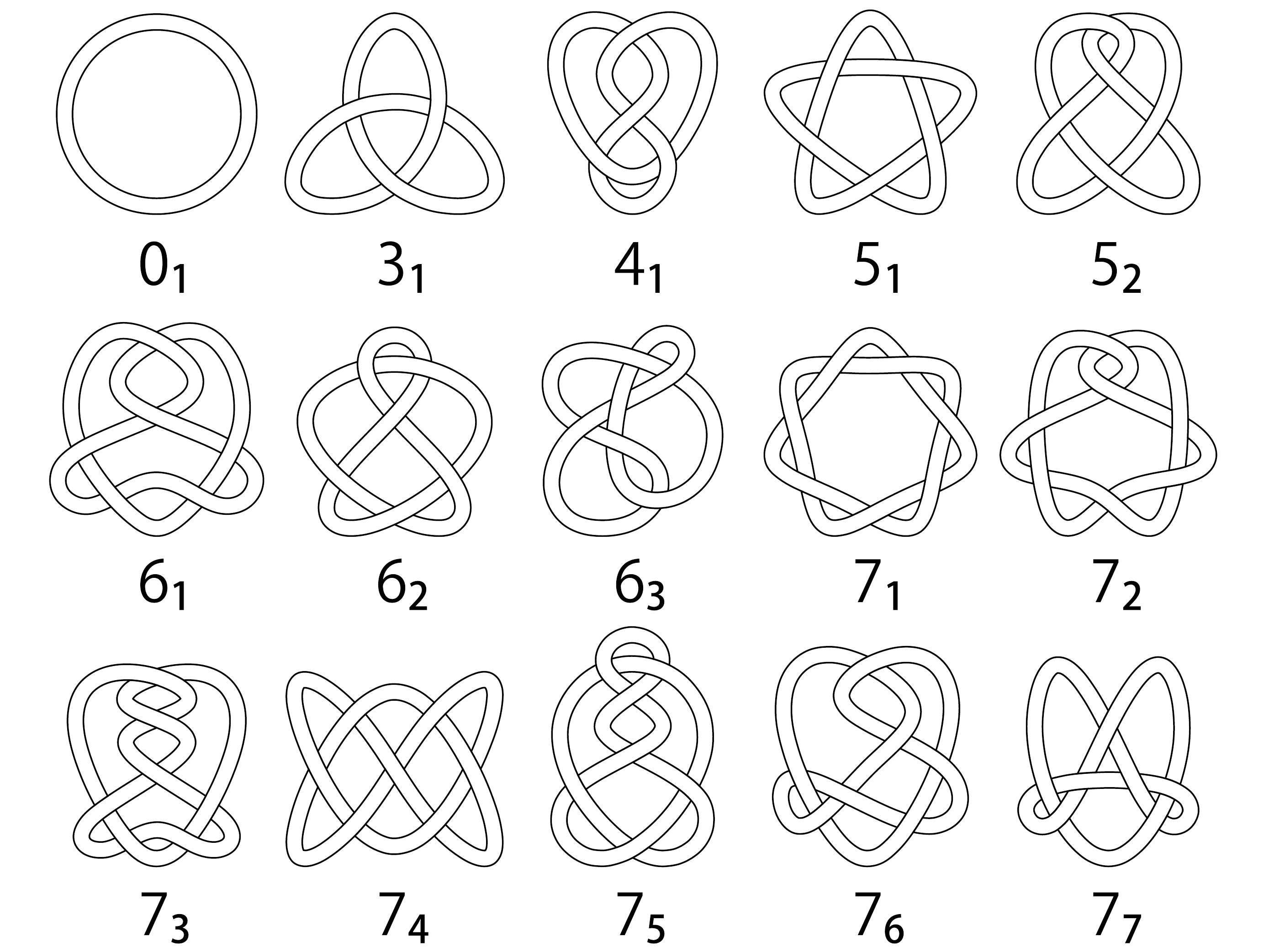}
  \caption{Unknot (the trivial knot, $0_1$)  and the prime knots with up to seven minimal crossings. }
  \label{fig0}
\end{center}
\end{figure}
%-----------------------------------------------------------------------

Let us explain important properties of the knotting probability. In a model of RP or SAP  of $N$ segments we denote the knotting probability of a knot $K$ by $P_K(N)$. 
It was shown that the knotting probabilities for the bead-rod model are well approximated  as a function of  $N$ by \cite{Deguchi-Tsurusaki1997}
%-----------------------------------------------------------------------
\begin{equation}
P_K(N)=C_K\tilde{N}^{m(K)}\exp(-\tilde{N}) \, , 
\label{eq:4formula}
\end{equation}
where $\tilde{N}$ is given by  
%-----------------------------------------------------------------------
\begin{equation}
\tilde{N}=\frac{N-\Delta N(K)}{N_K} \, .   
\label{eq:finite-size}
\end{equation}   
We call parameters $C_K$, $m(K)$, $N_K$ and $\Delta N(K)$
the knot coefficient (or coefficient), the knot exponent (or exponent),  the characteristic length and the finite-size correction 
of the knotting probability of knot $K$, respectively.  

We derive formula (\ref{eq:4formula}) by assuming  
the large-$N$ asymptotic expansion of the knotting probabilities. 
For simplicity, let us consider a model of lattice polygons. We denote by $Z_{K}(N)$ the number of lattice polygons of $N$ segments with a knot $K$ and by $Z_{All}(N)$ 
that of no topological constraint.  
The knotting probability of knot $K$ is given by $P_K(N)=Z_K(N)/Z_{All}(N)$.  
We assume that  for topological conditions $K$ including no constraint $All$
the numbers $Z_K(N)$ have the large-$N$ asymptotic expansion
\begin{equation}
\log Z_K(N) = \kappa_K N + m_K  \log N +  \log Z_K^{(0)} + O(1/N) . \label{eq:ZK}
\end{equation}  
By taking the exponential of Eq.  (\ref{eq:ZK}) and introducing parameters $N_K$ and $m(K)$  by 
$1/N_K = \kappa_{All}-\kappa_K$ and $m(K) = m_K-m_{All}$, respectively, we have 
\begin{equation} 
P_K(N)= C_K \left( {N} /{N_K}  \right)^{m(K)} \exp \left(- N/N_K \right) . \label{eq:3formula} 
\end{equation}
We call it the asymptotic formula of the knotting probability.  We obtain Eq. (\ref{eq:4formula})   
by replacing $N/N_K$ in  Eq. (\ref{eq:3formula}) with Eq. (\ref{eq:finite-size}). 

In several models of RP and SAP the estimates of $N_K$  for different knots 
are given by almost the same value as far as investigated.  
We therefore call it the characteristic length of the knotting probability\cite{JKTR}, 
and  denote it by $N_0$. For the cylindrical SAP  the $N_K$s are evaluated as the same for 145 knots  with respect to errors \cite{UD2015}. 

For a composite knot consisting of knots $K_1$ and $K_2$, denoted by $K_1 \# K_2$,  
its exponent and coefficient are approximately equal to  
the sum and the product of those of knots $K_1$ and $K_2$, respectively,  in several models of RP and SAP. We call such properties  factorization properties of exponents and coefficients, respectively.    
It was suggested that the factorization properties are derived from the local knot conjecture that the knotted region in a knotted SAP is localized \cite{Orlandini1998,Katritch00,Marcone}.

In the present paper we shall show that coefficients $C_K$ for almost all the prime knots are well approximated by exponentially decaying functions of radius $r_{\rm ex}$ in the cylindrical SAP model. We show it numerically for the prime knots with less than or equal to seven minimal crossings. Quite interestingly, only for the trefoil knot the coefficient $C_{3_1}$ increases with respect to radius $r_{\rm ex}$. It follows that the majority of nontrivial knots are given by the trefoil knot and its composite knots if radius $r_{\rm ex}$ is rather large such as $r_{\rm ex}=0.1$. 

The knot coefficients $C_K$ have an interesting property that the sum of the knot coefficients over all prime knots is given by 1. We call it the sum rule of knot coefficients for prime knots.  
We shall show that it is consistent with the formula of the knotting probability.  
We derive an infinite number of sum rules such that the sum of the knot coefficients 
over all composite knots consisting of $n$ prime knots is given by $1/n!$ 
for positive integers $n$.   We shall show  in section 5 that they are derived 
from the factorization properties of exponents and coefficients. Furthermore, we suggest that the sum rules give a numerical support for the local knot conjecture. 

In order to investigate how far the asymptotic behavior is important in the knotting probability as a function of segment number $N$ we apply the three-parameter asymptotic 
formula (\ref{eq:3formula}) to the data points of the knotting probability against segment number $N$.  The estimates of exponent $m(K)$ of a knot $K$ are much closer to some integers than in the case of the four-parameter formula (\ref{eq:4formula}), although the $\chi^2$ values of the fitted curves are larger than those of Eq. (\ref{eq:4formula}). 
It seems that the results of  the asymptotic formula  (\ref{eq:3formula}) is more similar to those of on-lattice SAP, where the estimates of the entropic exponent are given by integers  \cite{Orlandini1996,Orlandini1998}.  Here we remark that the entropic exponent corresponds to the exponent $m(K)$ of a knot $K$ in the notation of the present paper.

The contents of the paper consist of the following.  In section 2 we explain the algorithm for generating the cylindrical SAP with radius $r_{\rm ex}$ and then present some knot invariants by which we detect the knot type of a given polygon. We also give the numbers of polygons we have generated in the present research, which lead to the estimates of statistical errors. 
In section 3 we show that the four-parameter formula  (\ref{eq:4formula}) gives good fitted curves to the data points of the knotting probability versus segment number $N$. 
We exhibit  fitted  curves to the data of the knotting probability against $N$ 
for several  prime and composite knots.  
We then present fundamental and generic properties of the knotting probability as addressed briefly in Introduction. 
In section 4 we formulate important  properties of knot coefficients $C_K$. 
First, we argue that knot coefficient of a knot $K$ determines the maximum value of the knotting probability of the knot $K$. Second, we show numerically how the parameters $C_K$ 
for prime knots depend on the radius of cylindrical segments of the SAP. Third, 
we numerically confirm the factorization property of knot coefficients.   
In section 5 we show some important aspects of the knotting probability. 
We argue numerically that the results of the cylindrical SAP at $r_{\rm ex}=1/8$ correspond to those of lattice SAP. We derive the sum rules for knot coefficients  $C_K$ 
from Eq.  (\ref{eq:4formula}) of the knotting probability.  
We numerically confirm  the sum rule of coefficients $C_K$ for prime knots. 
We suggest that it gives a numerical support for  the local knot conjecture. 
In section 6 we discuss how effective the asymptotic expansion of the knotting probability is.  
In section 7 we give some concluding remarks.

%***********************************************************************
% Section: 2
% Numerical methods
%***********************************************************************
\section{Numerical methods}

Let us explain the method for evaluating the knotting probability of a given knot $K$ for the cylindrical SAP: We generate an ensemble of cylindrical SAP by the Monte-Carlo method, detect the knot type of the SAP by calculating some knot invariants, and then evaluate the knotting probability of the knot $K$ for the cylindrical SAP.

\subsection{Algorithm for generating cylindrical SAP}
We construct an ensemble of SAP consisting of $N$ cylindrical segments with radius $r_{\rm ex}$ as follows \cite{UD2015}. First, we construct an initial polygon by an equilateral regular $N$-gon, where the vertices have numbers from $1$ to $N$, consecutively. Second, we choose two vertices randomly out of the $N$ vertices. Suppose that they are given by numbers $p_1$ and $p_2$.We rotate a sub-chain between the vertices $p_1$ and $p_2$ around the straight line connecting them by an angle chosen randomly from $0$ to $2\pi$. Third, we check whether the rotated sub-chain has any overlap with the other part of the polygon or not. If the distance between every pair of non-neighboring segments (or polygonal edges) of the polygon is larger than $2 r_{\rm ex}$, we find that the polygon does not have any overlap. If it has no overlap, we employ the rotated configuration as the cylindrical SAP in the next Monte-Carlo step. If it has an overlap, we employ the previous configuration of SAP before rotation in the next Monte-Carlo step. Then, we repeat this procedure many times such as $2N$ times.

In the case of $r_{\rm ex}=0$, the SAPs generated by the  above  algorithm are given by equilateral random polygons. The algorithm for generating cylindrical SAPs with $r_{\rm ex}=0$ is also called the polygonal folding method (PFM) \cite{Millett1994}. The ergodicity of PFM is shown in Ref.  \cite{Millett1994,Kapovich1996} (see also \cite{Millett2011}).

\subsection{Method for evaluating the knotting probability}
We detect the knot type of a given SAP by evaluating mainly the values of the two knot invariants: The absolute value of the Alexander polynomial $|\Delta_K(t)|$ evaluated at $t=-1$ and the Vassiliev invariant of the second order $v_2(K)$ for a knot $K$. If a given SAP has the same values of two knot invariants as a knot $K$, we assume that the topology of the polygon is given by the knot $K$. For some cases we also evaluate the Vassiliev invariant of the third order, as we shall see later. 

Some pairs of knots have the same values of the two knot invariants in common.
For example, both knot $7_4$ and knot $3_1\# 5_1$ have the same values $|\Delta_K(-1)|=15$ and  $v_2(K)= 4$. Therefore, we cannot distinguish between knot $7_4$ and knot $3_1\sharp5_1$ only by calculating the two knot invariants. In order to distinguish them we evaluate the Vassiliev invariants of the third order for such polygons. 

The Vassiliev invariants of any order can be calculated by the method of the quasi-classical expansion of the $R$-matrix of the quantum group \cite{Deguchi-Tsurusaki-PLA,JKTR}. However, we employ the algorithm due to Polyak and Viro to calculate the Vassiliev invariants of the second order and the third order \cite{Polyak-Viro}.  In fact, by the latter method the Vassiliev invariants are calculated only through the Gauss codes \cite{Murasugi} (or the Dowker codes).

\subsection{Number of SAPs generated in the simulation}

In the present simulation for each value of radius $r_{\rm ex}$ 
we generated $2\times10^5$ polygons for $N \le 4, 000$,  $10^5$ polygons for $N$ satisfying $4, 000 \le N \le 6, 000$,  $5 \times10^4$ polygons for $N$ satisfying $6, 000 \le N \le 8, 000$, and $4 \times10^4$ polygons for $N$ satisfying $8, 000 \le N \le10, 000$.

In the present simulation the number of segments $N$ is given from $100$ to $3, 000$ for the cylindrical SAP of zero thickness ($r_{\rm ex}= 0$), i.e. equilateral random polygons; from $100$ to $3000$ for cylindrical SAP with $r_{\rm ex}=0.005$ and $0.01$;  from $100$ to $4000$ with $r_{\rm ex}=0.02$; from $100$ to $5, 000$ with $r_{\rm ex}=0.03$; from $100$ to $7, 000$  with $r_{\rm ex}=0.04$;  from $100$ to $8, 000$ with $r_{\rm ex}=0.05$; from $100$ to $10^4$ with $r_{\rm ex}=0.06$, $0.08$ and $0.1$.

*********************
% Section: 3
% knotting probabilities of various knots 
%***********************************************************************
\section{Knotting probabilities of various knots}

%***********************************************************************
% Subsection: 3.1 
% knotting probabilities of trivial & 3_1 & 4_1 & 5_1 & 5_2 knot
%***********************************************************************
\subsection{Knotting probability for prime knots}

%-----------------------------------------------------------------------
% Figure 2. Knot probability of 31 knot
\begin{figure}[htbp]
\begin{center}
\includegraphics[width=0.95\hsize]{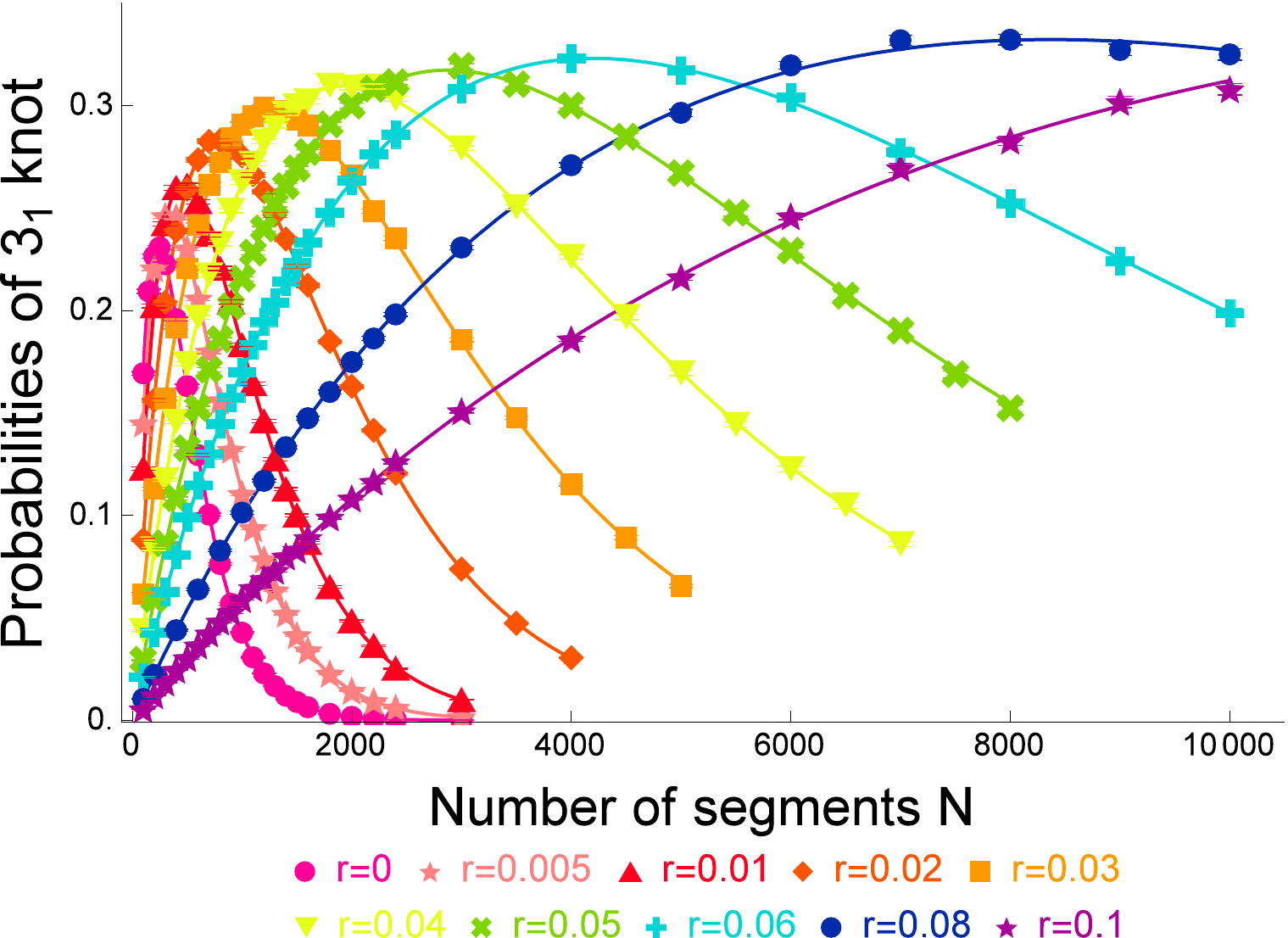} 
 \caption{Knotting probability of the trefoil knot ($3_1$) versus the number of segments $N$ for the cylindrical SAPs with ten different values of the cylindrical radius. The plots for the values of radius $r_{\rm ex}$ given by 0.0, 0.005, 0.01, 0.02, 0.03, 0.04, 0.05, 0.06, 0.08 and 0.10, are depicted by circles (red), stars, upper triangles, diamonds, squares, lower triangles, saltires or Xs, crosses, circles (blue) and stars (purples), respectively.  The fitted curves have the best estimates of the parameters of Eq. (\ref{eq:4formula}) listed in Table \ref{tab:31}. } 
  \label{fig:31}
\end{center}
\end{figure}

%-----------------------------------------------------------------------
%
% Table 1
%
\begin{table*}[htbp] 
\begin{center} 
\begin{tabular}{c|ccccc} 
\hline 
$r_{\rm ex}$ & $C_K$ & $m(K)$  & $N_K$ & $\Delta N(K)$ & $\chi^2$/DF \\
\hline 
 0 & $0.6183 \pm 0.0012$ & $0.852 \pm 0.014$ & $257.1 \pm 1.2$ & $18.4 \pm 2.2$ & 
$0.56$  \\ 
0.005 & $0.6643 \pm 0.0018$ & $0.832 \pm 0.015$ & $380.5 \pm 2.3$ & $22. \pm 
2.7$ & $1.14$  \\ 
0.01 & $0.7039 \pm 0.0017$ & $0.856 \pm 0.013$ & $517.7 \pm 3.1$ & $19. \pm 
2.5$ & $1.19$  \\ 
0.02 & $0.7644 \pm 0.0010$ & $0.8929 \pm 0.0067$ & $867.7 \pm 3.3$ & $13.9 \pm 
1.7$ & $0.68$  \\ 
0.03 & $0.8087 \pm 0.0013$ & $0.9303 \pm 0.0074$ & $1348.3 \pm 6.9$ & $9.1 \pm 
2.2$ & $1.13$  \\ 
0.04 & $0.8404 \pm 0.0014$ & $0.9414 \pm 0.0064$ & $2047. \pm 11.$ & $6.7 \pm 
2.3$ & $1.15$  \\ 
0.05 & $0.8607 \pm 0.0013$ & $0.9369 \pm 0.0051$ & $3051. \pm 17.$ & $13.7 \pm 
2.0 $ & $0.91$  \\ 
0.06 & $0.8765 \pm 0.0011$ & $0.9385 \pm 0.0034$ & $4455. \pm 21.$ & $14.2 \pm 
1.5$ & $0.41$  \\ 
0.08 & $0.902 \pm 0.0031$ & $0.9513 \pm 0.0069$ & $8770. \pm 140.$ & $18.4 \pm 
2.9$ & $0.94$  \\ 
0.1 & $0.926 \pm 0.0120$ & $0.9520 \pm 0.0076$ & $16690. \pm 550.$ & $24.5 \pm 
3.5$ & $1.01$  \\ 
\hline 
\end{tabular} 
\end{center} 
\caption{Best estimates of the parameters in Eq. (\ref{eq:4formula}) for the knotting probability of the trefoil knot $3_1$ with ten different values of cylindrical radius $r_{\rm ex}$. }
\label{tab:31}
\end{table*} 
%-----------------------------------------------------------------------

\subsubsection{Maximum probability of trefoil knot increases  as the excluded volume of SAP increases}

Let us denote  by  the symbol $P_K(N, r_{\rm ex})$ 
the knotting probability of a knot $K$ for the cylindrical SAP consisting of  $N$ cylindrical segments with radius $r_{\rm ex}$. 

In Fig. \ref{fig:31}  the knotting probabilities of the trefoil knot ($3_1$) for the cylindrical SAP with radius $r_{\rm ex}$ are plotted against segment number $N$ for various values of cylindrical radius $r_{\rm ex}$.  Here we have plotted them for ten different values of cylindrical radius such as $r_{\rm ex}= 0.0, 0.005, 0.01, 0.02 \cdots,$ and so on.   

We observe in Fig. \ref{fig:31} that  the maximum value of the knotting probability of knot $3_1$  increases as radius $r_{\rm ex}$ increases. 
The peak height of each plot  increases gradually as radius $r_{\rm ex}$ increases, while 
the peak position, i.e. the number of segments $N$ at which the knotting probability gives the maximum value, is shifted to the right as radius $r_{\rm ex}$ increases. 
The peak position is approximately given by the characteristic length $N_{3_1}$ if we assume eq. (\ref{eq:4formula}).  Here we remark that the exponent of trefoil knot, $m(3_1)$, is estimated 
as roughly equal to 1.0, as shown later. 

\begin{figure}[htbp]
\begin{center}
\includegraphics[width=0.95\hsize]{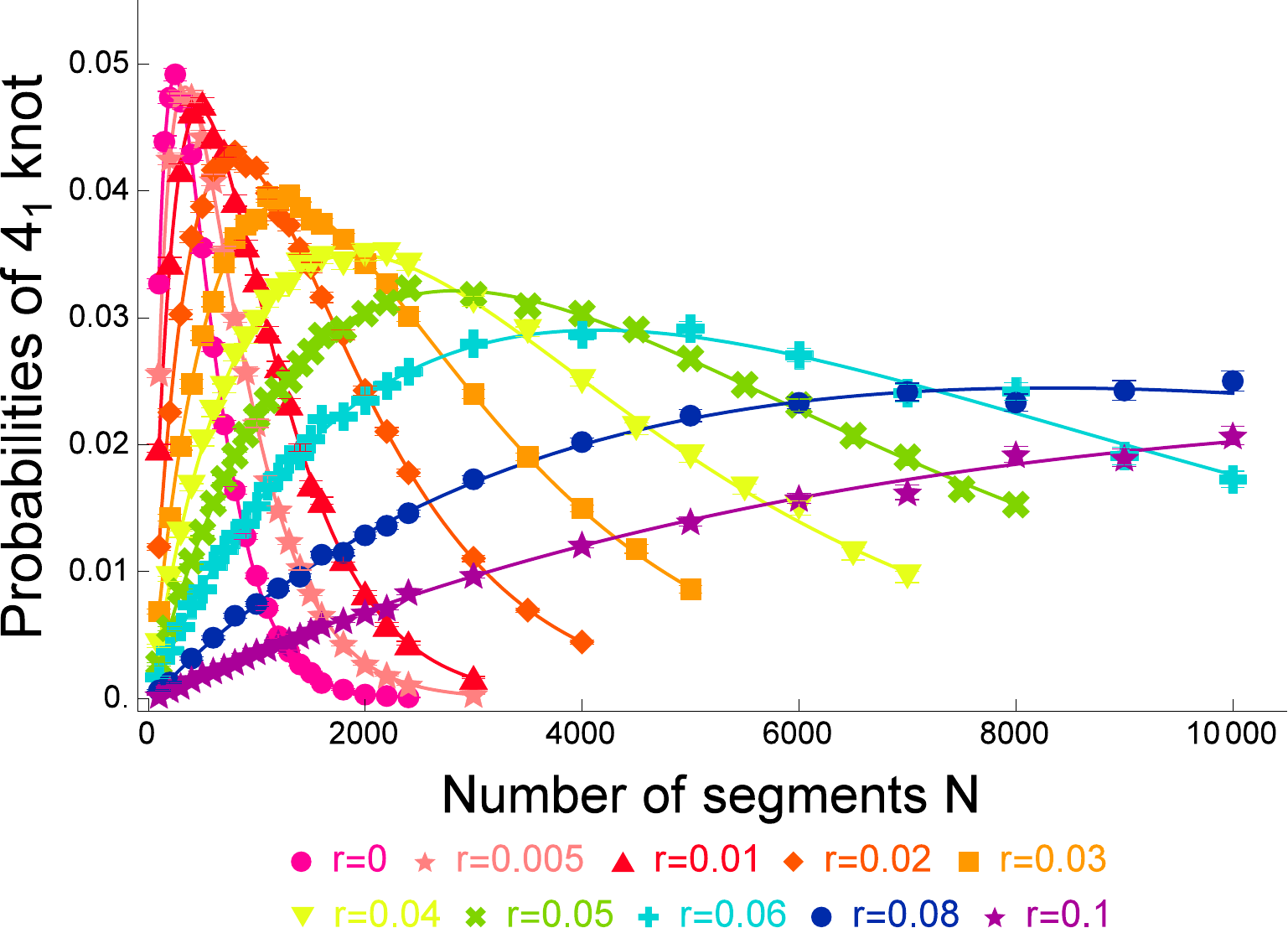}
  \caption{Knotting probability of the figure-eight knot $4_1$ for cylindrical SAPs with ten different values of the cylindrical radius: $r_{\rm ex}=0.0$, 0.005, 0.01, 0.02, 0.03, 0.04, 0.05, 0.06, 0.08 and 0.1. The fitted curves have the best estimates of the parameters of Eq. (\ref{eq:4formula}) listed in Table \ref{tab:41}. } 
 \label{fig:41}
\end{center}
\end{figure}

In Fig. \ref{fig:41} the knotting probabilities of the figure-eight knot ($4_1$) for the cylindrical SAP with radius $r_{\rm ex}$ are plotted against segment number $N$ for various  values of radius $r_{\rm ex}$. In Fig. \ref{fig:41}  the maximum value of the knotting probability of knot $4_1$ decreases with respect to radius $r_{\rm ex}$.    

The fitted curves in Figs. \ref{fig:31} and \ref{fig:41}  are given by formula (\ref{eq:4formula}). 
They are good, since the $\chi^2$ values are less than 2.0 for all the curves. 
Here we remark that the best estimates of parameters of Eq. (\ref{eq:4formula}) are listed in 
Tables and \ref{tab:31} and \ref{tab:41} for knots $3_1$ and $4_1$, respectively,  
together with the $\chi^2$ value per degree of freedom (DF).

%----------------------------------------------------------------
% Figure 3. knotting probabilities of 41 knot
%
%-----------------------------------------------------------------------

The knotting probabilities of such prime knots that have less than or equal to seven minimal crossings for the cylindrical SAP with radius $r_{\rm ex}$ are fitted  by formula (\ref{eq:4formula}). 
The best estimates of the parameters of Eq. (\ref{eq:4formula}) 
together with the $\chi^2$ value per DF are listed in 
Tables \ref{tab:56} and \ref{tab:7} in Appendix A:  for knots $5_1$, $5_2$, $6_1$, $6_2$, and $6_3$ in Table \ref{tab:56};  for $7_1$, $7_2$, $7_3$  $7_4$, $7_5$, $7_6$  and $7_7$ in Tables \ref{tab:7}.  
The $\chi^2$ values are smaller than 2.0 for all the fitted curves.

We can show that the maximum value of the knotting probability of a knot $K$ is determined by the coefficient $C_K$.  We shall show it in section \ref{sec:maxP} by making use of  Eq. (\ref{eq:4formula}).

The increase of  the maximum value of the knotting probability 
for a nontrivial knot with respect to  the excluded volume should be interesting and not trivial.  It is only the case for the trefoil knot ($3_1$) among all the prime knots as far as we have investigated. For other prime knots the maximum value of the knotting probability decreases as radius $r_{\rm ex}$ increases. We confirm it 
from the estimates of knot coefficients $C_K$ 
for knots $4_1$ in Table \ref{tab:41},  
 $5_1$, $5_2$, $6_1$, $6_2$, and $6_3$ in Tables \ref{tab:56},  
 $7_1$, $7_2$, $7_3$  $7_4$, $7_5$, $7_6$  and $7_7$ 
in Table \ref{tab:7} in Appendix A.

We observe in Figs. \ref{fig:31} and \ref{fig:41} 
as well as in those of knots $5_1$ and $5_2$  that the maximum value of the knotting probability for knot $K$ decreases as the minimal crossing number of the knot $K$ increases at least among the four prime knots, $3_1$, $4_1$, $5_1$ and $5_2$.

\subsubsection{Fitted curves with knot exponent close to 1}

The estimate of exponent $m(K)$ of a knot $K$ is roughly given by  1.0 for  the four prime knots, $3_1$, $4_1$, $5_1$ and $5_2$. In Table \ref{tab:31} the exponent $m(3_1)$ of the trefoil knot is approximately given by 1.0 .  However, if we consider the estimates of errors,  it is clearly smaller than 1.0  with respect to errors. 
Here we recall that 
the best estimates of parameters of eq. (\ref{eq:4formula}) are listed in 
Tables \ref{tab:31} and \ref{tab:41} together with the $\chi^2$ value per DF for knots 
$3_1$ and $4_1$, respectively.

\subsubsection{Small-$N$ region}

When segment number $N$ is small  such as much smaller than the characteristic length $N_K$,  
the knotting probability of a nontrivial knot $K$ can be approximated by a straight line as a function of $N$.  We observe it in Figs. \ref{fig:31} and \ref{fig:41} for knots $3_1$ and $4_1$, respectively. 
We also see it for knots $5_1$ and $5_2$.   

In the small-$N$ region if we fix a number of segments $N$ the knotting probability $P_K(N, r_{\rm ex})$ decreases with respect to radius $r_{\rm ex}$, as shown in  Figs. \ref{fig:31} and \ref{fig:41}. If we assume the four-parameter formula (\ref{eq:4formula}) it is a consequence of the fact that  the characteristic  length increases rapidly with respect to the excluded-volume parameter, i.e. the cylindrical radius $r_{\rm ex}$. For small-$N$ region such as $N \ll N_K$, formula (\ref{eq:4formula}) is approximated by a linear function of $N$ as 
\begin{equation} 
P_K(N, r_{\rm ex}) \approx C_K {\frac {N-\Delta N(K)}{N_K}} \, . 
\label{eq:small-N}
\end{equation}
Here for simplicity we have assumed that the exponent $m(K)$ of knot $K$ is almost given by 1 if 
knot $K$ is a prime knot.  As radius $r_{\rm ex}$ increases the characteristic length $N_K$ increases rapidly while the coefficient $C_K$ does not change very much, so that the knotting probability decreases with respect to $r_{\rm ex}$ for a given fixed number of segments $N$.

% -------------------------------------------------------------
%
\begin{table*}[htbp]
\begin{center} 
\begin{tabular}{c|ccccc} 
\hline 
$r_{\rm ex}$ & $C_K$ & $m(K)$  & $N_K$ & $\Delta N(K)$ & $\chi^2$/DF \\
\hline 
0 & $0.13113 \pm 0.00090$ & $0.841 \pm 0.041$ & $257.5 \pm 3.8$ & $31.9 \pm 6.0$ & 
$1.43$  \\ 
0.005 & $0.1279 \pm 0.00078$ & $0.793 \pm 0.027$ & $383.5 \pm 4.5$ & $38.0 \pm 
4.4$ & $0.94$  \\ 
0.01 & $0.12575 \pm 0.00062$ & $0.929 \pm 0.033$ & $492.8 \pm 7.2$ & $20.3 \pm 
6.0$ & $1.36$  \\ 
0.02 & $0.11522 \pm 0.00047$ & $0.887 \pm 0.018$ & $860.8 \pm 9.3$ & $26.7 \pm 
4.2$ & $0.89$  \\ 
0.03 & $0.10542 \pm 0.00040$ & $0.902 \pm 0.014$ & $1355. \pm 15.$ & $31.4 \pm 
3.8$ & $0.75$  \\ 
0.04 & $0.09443 \pm 0.00046$ & $0.891 \pm 0.015$ & $2096. \pm 28.$ & $34.3 \pm 
4.5$ & $0.96$  \\ 
0.05 & $0.08711 \pm 0.00033$ & $0.937 \pm 0.012$ & $3004. \pm 39.$ & $26.8 \pm 
4.4$ & $0.59$  \\ 
0.06 & $0.07865 \pm 0.00058$ & $0.932 \pm 0.019$ & $4420. \pm 120.$ & $27.5 \pm 
7.7$ & $1.34$  \\ 
0.08 & $0.0663 \pm 0.0011$ & $0.938 \pm 0.031$ & $8820. \pm 670.$ & $46. \pm 
11.$ & $1.82$  \\ 
0.1 & $0.0595 \pm 0.0025$ & $0.981 \pm 0.024$ & $15700. \pm 1600.$ & $39. \pm 
10.$ & $0.74$  \\ 
\hline 
\end{tabular} 
\end{center} 
\caption{Best estimates of the parameters in Eq. (\ref{eq:4formula}) for the knotting probability of the figure-eight knot $4_1$ with ten different values of cylindrical radius $r_{\rm ex}$. }
\label{tab:41}
\end{table*}

Here we remark that the finite-size corrections $\Delta N(K)$ increase slightly as cylindrical radius $r_{\rm ex}$ increases.

\subsubsection{Knots with the same crossing number}

For the knotting probabilities of knots $5_1$ and $5_2$ 
that of knot $5_2$ is almost twice as large as that of knot $5_1$, 
although they have the same  minimal crossing numbers.  
We can confirm it from the estimates of knot coefficients $C_K$ 
listed in Table \ref{tab:56} of Appendix A. 
Here we remark that knot $5_1$ is a torus knot, while  
knot $5_2$ is a twsit  knot \cite{Murasugi}. 

Interestingly, it is also the case for knots $7_1$ and $7_2$. 
The knot coefficient of knot $7_2$, which is a twist knot, 
is more than twice as large as that of knot $7_1$, 
which is a torus knot, as listed in Table \ref{tab:7} of Appendix A.

%***********************************************************************
% Section 3.2
%
% knotting probabilities of composite knot
%***********************************************************************
\subsection{Knotting probabilities of composite knots}

Let us introduce prime knots and composite knots \cite{Murasugi}. 
If a diagram of a knot $K$ is decomposed into two diagrams of nontrivial knots $K_1$ and $K_2$ by cutting two points in the diagram of $K$, we say that it is composed of the two knots and denote it by  $K=K_1 \# K_2$. We also say that it is the product of them. 
If a knot $K$ cannot be decomposed into a product of two nontrivial knots, we say that it is prime.  

\subsubsection{Factorization properties of exponents and coefficients}

By applying Eq. (\ref{eq:4formula}) to the data points of the knotting probability versus segment number $N$ we observe that the best estimate of the exponent $m(K)$ for a composite knot $K= K_1 \# K_2$ is given by the sum of the best estimates of the exponents for constituent knots $K_1$ and $K_2$ \cite{JKTR}   
\begin{equation}
m(K_1 \# K_2) = m(K_1) + m(K_2)  \, .  \label{eq:factor0}
\end{equation}
We call it the factorization property of exponents $m(K)$. 
An analytical derivation of Eq. (\ref{eq:factor0}) was argued by assuming the local knot picture \cite{TD95}. 
In several models of RP and SAP the estimate of exponent $m(K)$ of a composite knot is approximately given by the number of prime knots of which the composite knot consists.  

Similarly, in several models of RP and SAP 
we observe the factorization property of knot coefficients $C_K$:  
for a composite knot $K_1 \# K_2$ consisting of 
knots $K_1$ and $K_2$  we have 
%\begin{equation}
$C_{K_1 \# K_2} = C_{K_1} C_{K_2}$,  if $K_1 \ne K_2$,    
%\end{equation}  
among the estimates of coefficients $C_{K_1 \# K_2}$, $C_{K_1}$ and $C_{K_2}$.  
For a composite knot $K$ consisting of $n$ prime knots such that  
there are $n_j$ copies of prime knots $K_j$ and 
the sum of integers $n_j$ is given by $n$, we have
\begin{equation}
C_K = \prod_j \left( C_{K_j} \right)^{n_j} / n_j ! \, .  \label{eq:fact3}
\end{equation}
We remark that in the present research we shall numerically show the factorization property 
for the fitting parameters $C_K$ of the four-parameter formula  (\ref{eq:4formula}) 
where finite-size corrections $\Delta N(K)$ are taken into account.

For lattice knots the factorization property among coefficients $C_K$  for large $N$ is studied numerically by making use of the asymptotic expansion of the knotting probability \cite{Stella},  
which corresponds to Eq. (\ref{eq:3formula}) where no finite-size corrections $\Delta N(K)$ 
are considered.

It has been suggested that the factorization properties of exponents $m(K)$ and coefficients $C_K$ are favorable to the local-knot picture. Here we recall the local knot conjecture that for a RP or SAP with a nontrivial knot, the knotted part of the RP or SAP is localized in some way \cite{Orlandini1998,Katritch00,Marcone}.  
However, it is not trivial to show the suggestion even numerically \cite{Tubiana}.

\subsubsection{Fitted curves for composite knot of two trefoil knots }

%-----------------------------------------------------------------------
% Figure 4. knotting probabilities of 31#31 knot
%
\begin{figure}[htbp]
\begin{center}
  \includegraphics[width=0.9\hsize]{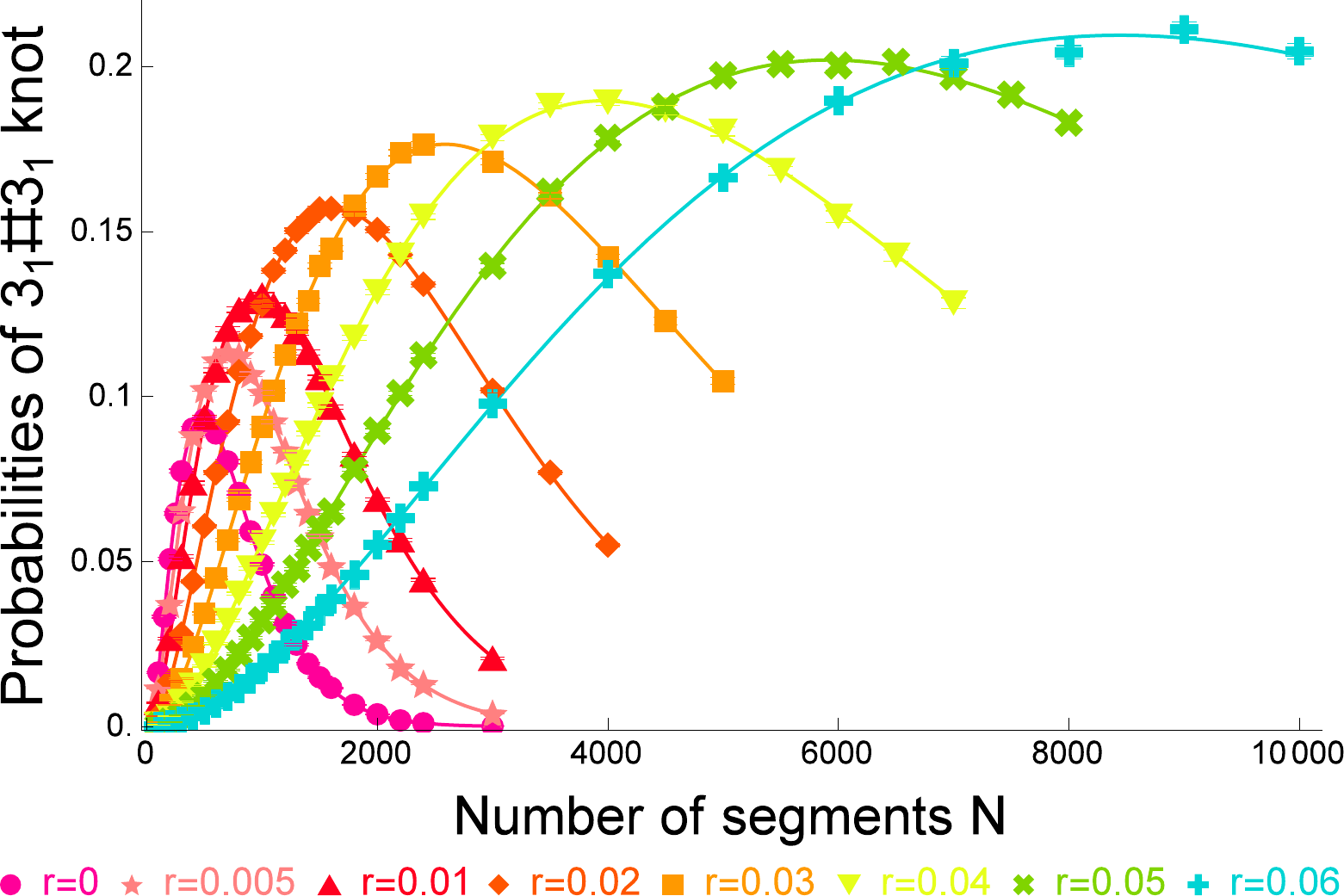}
  \caption{Knotting probability of composite knot $3_1 \# 3_1$ versus the number of segments $N$ for the cylindrical SAP with ten different values of radius: $r_{\rm ex}=0.0$, 0.005, 0.01, 0.02, 0.03, 0.04, 0.05, 0.06, 0.08 and 0.1. 
The fitted curves are given by applying Eq. (\ref{eq:4formula}), where the best estimates of the parameters are listed in Table \ref{tab:3131}.  }
  \label{fig:3131}
\end{center}
\end{figure}
%-----------------------------------------------------------------------

In Fig. \ref{fig:3131}, the numerical estimates of the knotting probability of composite knot $3_1\# 3_1$ are plotted against the number of segments $N$ for the cylindrical SAPs with ten different values of cylindrical radius $r_{\rm ex}$. Here we remark that the composite knot $3_1\# 3_1$ consists of the product of two trefoil knots $3_1$ and $3_1$. The fitted curves are given by eq. (\ref{eq:4formula}). The fitted curves fit very well to the data points. The $\chi^2$ values are small.  Here, the best estimates of the fitting parameters and the $\chi^2$ values are given in Table \ref{tab:3131}.

% -----------------------------------
% Table 3 
%
\begin{table*}[htbp] 
\begin{center}
\begin{tabular}{c|ccccc} 
\hline 
$r_{\rm ex}$ & $C_K$ & $m(K)$  & $N_K$ & $\Delta N(K)$ & $\chi^2$/DF \\
\hline 
0 & $0.2010 \pm 0.0021$ & $1.759 \pm 0.018$ & $261.1 \pm 1.3$ & $26.3 \pm 1.8$ & 
$0.61$  \\ 
0.005 & $0.2367 \pm 0.0033$ & $1.817 \pm 0.024$ & $380.5 \pm 2.8$ & $19.3 \pm 
2.7$ & $1.19$  \\ 
0.01 & $0.2740 \pm 0.0031$ & $1.801 \pm 0.020$ & $522.2 \pm 3.9$ & $24.8 \pm 2.7$ & 
$1.26$  \\ 
0.02 & $0.3212 \pm 0.0017$ & $1.8430 \pm 0.0098$ & $873.0 \pm 3.7$ & $24.6 \pm 
1.8$ & $0.47$  \\ 
0.03 & $0.3491 \pm 0.0033$ & $1.897 \pm 0.016$ & $1354. \pm 12.$ & $21.1 \pm 
3.7$ & $1.28$  \\ 
0.04 & $0.3744 \pm 0.0028$ & $1.899 \pm 0.012$ & $2069. \pm 16.$ & $22.4 \pm 
3.6$ & $0.87$  \\ 
0.05 & $0.3989 \pm 0.0034$ & $1.899 \pm 0.013$ & $3099. \pm 31.$ & $33.3 \pm 
4.4$ & $0.86$  \\ 
0.06 & $0.4028 \pm 0.0057$ & $1.940 \pm 0.017$ & $4337. \pm 70.$ & $31.1 \pm 
6.8$ & $1.26$  \\ 
\hline 
\end{tabular} 
\end{center} 
\caption{Best estimates of the parameters of Eq. (\ref{eq:4formula}) for the knotting probability of composite knot $3_1 \# 3_1$ with ten different values of cylindrical radius $r_{\rm ex}$. }
\label{tab:3131}
\end{table*}

It is clear in Fig. \ref{fig:3131} that the maximum value of the knotting probability of the composite knot $3_1\# 3_1$ increases as cylindrical radius $r_{\rm ex}$ increases. The peak position is approximately given by twice the characteristic length $N_0$. It is compatible with the fact that the estimate of the exponent of the composite knot is approximately given by 2.0: $m(3_1 \# 3_1) =2.0$ . Here we recall that the characteristic length $N_0$ also increases with respect to the cylindrical radius.

The numerical data of the knotting probability of a composite knot $3_1\# 3_1 \# 3_1$ are plotted against the number of segments $N$ for the cylindrical SAP with eight different values of cylindrical radius $r_{\rm ex}$ in Fig. \ref{fig:313131} of Appendix A. The fitted
 curves  given by eq. (\ref{eq:4formula}) are good. 
The best estimates are listed in Table \ref{tab:313131} of Appendix A.

We suggest that the maximum value of the knotting probability of the trefoil knot $3_1$ and that of a composite knot consisting of only the trefoil knot $3_1$ increases as the excluded-volume parameter $r_{\rm ex}$ increases, while the maximum value of the knotting probability of any prime knot other than knot $3_1$ decreases exponentially with respect to cylindrical radius $r_{\rm ex}$.

\subsubsection{Fitted curves for other composite knots}

\begin{figure}[htbp]
\begin{center}
  \includegraphics[width=1.0\hsize]{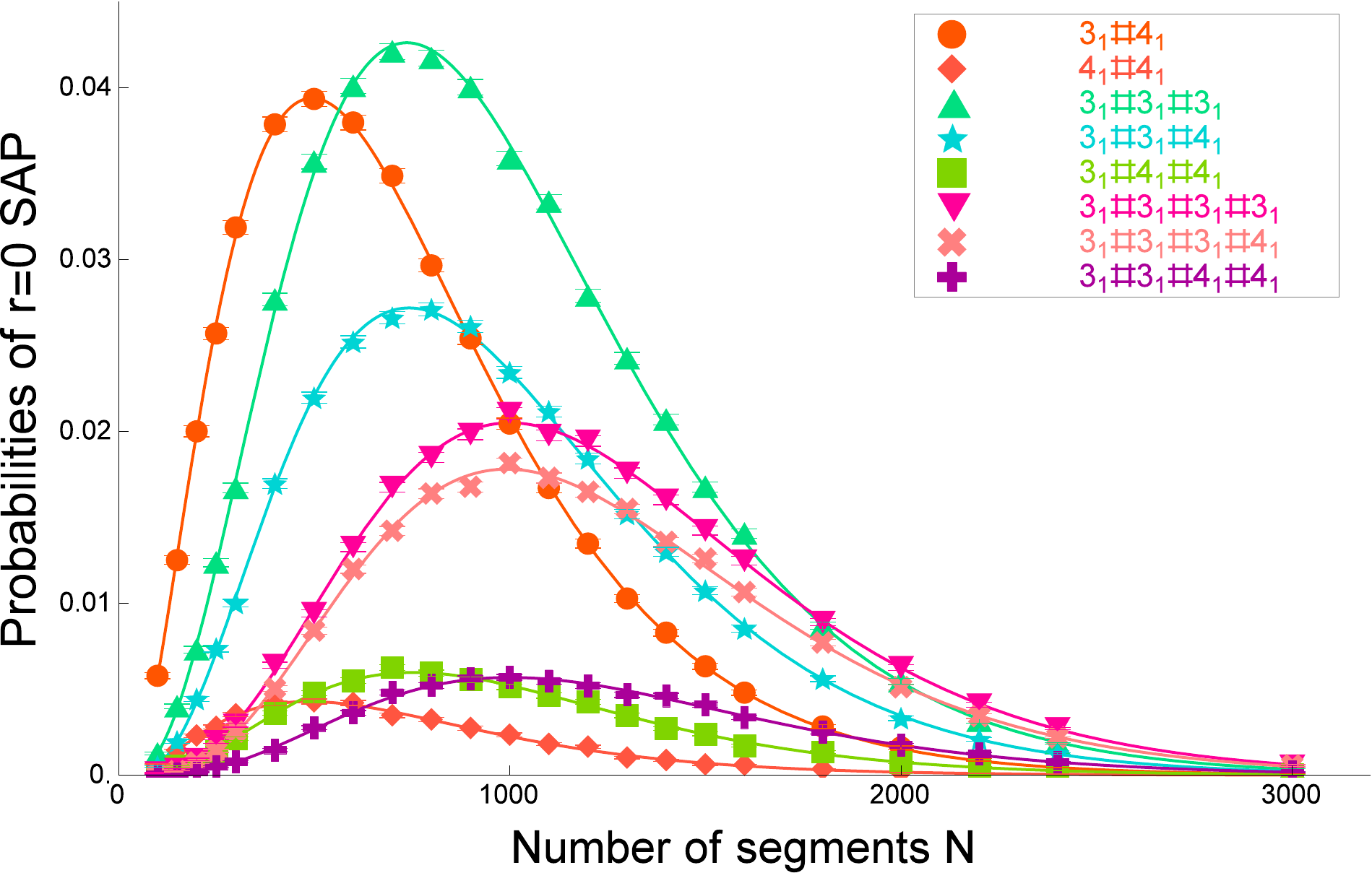}
\caption{ Knotting probabilities of several composite knots that contain 
the trefoil knot $3_1$ and the figure-eight knot $4_1$ 
  for the cylindrical SAP with zero thickness $r_{\rm ex}=0$, i.e., equilateral random polygons. 
Here the data points of composite knots $3_1 \# 4_1$, $4_1 \# 4_1$, $3_1 \# 3_1 \# 3_1$, 
  $3_1 \# 3_1 \# 4_1$,  $3_1 \# 4_1 \# 4_1$, $3_1 \# 3_1 \# 3_1 \# 3_1$, 
 $3_1 \# 3_1 \# 3_1 \# 4_1$, $3_1 \# 3_1 \# 4_1 \# 4_1$ are depicted by 
circles,  diamonds, upper triangles, stars, squares, lower triangles, saltires or Xs and crosses, respectively. The fitted curves are given by Eq. (\ref{eq:4formula}). }
  \label{fig:composite}
\end{center}
\end{figure}

In Fig. \ref{fig:composite} the knotting probabilities of various composite knots consisting of knots $3_1$ and $4_1$ such as $3_1 \# 4_1$, $3_1 \# 3_1 \# 4_1$ etc., are plotted against the number of segments $N$ for the cylindrical SAP in the case of zero thickness, i.e. for $r_{\rm ex}$=0. 
The fitted curves given by (\ref{eq:4formula}) 
are good and have small $\chi^2$ values per DF. 
The peak positions in Fig. \ref{fig:composite}  
are classified to the three types: those of composite knots consisting of two prime knots such as $3_1 \# 4_1$,   those of composite knots consisting of three prime knots such as $3_1 \# 3_1 \# 4_1$, and  those of composite knots consisting of four prime knots such as 
$3_1 \# 3_1 \# 3_1 \# 4_1$.

We have plotted the estimates of the knotting probabilities of composite knots against the number of segments $N$ for a large number of composite knots such as 130 composite knots \cite{UD2015}.  We observe the factorization properties of knot exponents and knot coefficients 
such as given in Eq. (\ref{eq:factor0}) and Eq. (\ref{eq:fact3}), respectively.

% -----------------------------------------------------------------
% Figure 5. knotting probabilities of many composite knots
%
%-----------------------------------------------------------------------
%
% Sec 4
%
\section{Fundamental properties of knot coefficients} 

%***********************************************************************
% Subsection: 4.1  
% Maximum value of knotting probabilities 
%
%***********************************************************************
\subsection{Knot coefficients determine the maximum of knotting probability}
\label{sec:maxP}

We now show that the knot coefficients $C_K$ mainly determine the maximum value of 
the knotting probability of a knot $K$.   

We recall that the estimates of exponents $m(K)$ for several prime knots are close to 1.0 but not equal to 1.0 with respect to errors.  Here we also recall that  the best estimates of the exponents $m(K)$  defined in Eq. (\ref{eq:4formula})  for knots $3_1$ and $4_1$ are given in Tables  \ref{tab:31} and \ref{tab:41}, 
respectively,  
and for knots with five and six minimal crossings and those of seven minimal crossings 
are listed in Tables  \ref{tab:56} and \ref{tab:7}, respectively, in Appendix A.

The range of the best estimates of exponent $m(K)$ is given from 0.8 to 1.2 for the prime knots. The best estimate of $m(K)$ increases slightly as cylindrical radius $r_{\rm ex}$ increases.

Let us express the maximum values of the knotting probability in terms of the fitting parameters of Eq. (\ref{eq:4formula}).   By taking its derivative with respect to $N$ we have 
%-----------------------------------------------------------------------
% Derivative of knotting probabilities of Nontrivial knot
\begin{equation}
\frac{dP(N,r_{\rm ex},K)}{dN}=\frac{C_K}{N_K}\tilde{N}^{m(K)-1}(m(K)-\tilde{N})\exp(-\tilde{N})
\label{Eq005}
\end{equation}
%-----------------------------------------------------------------------
Therefore,  the knot probability of a nontrivial knot $K$ 
has the maximum value at $\tilde{N}=m(K)$:  
\begin{equation}
N = m(K) N_K + \Delta N(K) .
\end{equation}
The maximum value of the knotting probability of knot $K$ is thus given by 
%-----------------------------------------------------------------------
% Peak height of knotting probabilities of Nontrivial knot
\begin{equation}
{\rm Max Prob}(K)    
= C_K m(K)^{m(K)}\exp(-m(K)) \, . 
\label{Eq006}
\end{equation}
%-----------------------------------------------------------------------

The value of $m(K)^{m(K)} \exp[-m(K)]$ does not change 
very much  when  the value of exponent $m(K)$ varies from $0.8$ to $1.2$.
In fact, the value of $m(K)^{m(K)}\exp(-m(K))$ is given by $0.376$ for  $m(K)=0.8$, and $0.368$ for $m(K)=1$.  Therefore, the maximum value of the knotting probability of a given prime knot $K$ depends mainly  on the coefficient $C_K$.

%%%%%%%%%%%%%%%%%%%%%%%%%%%%%%%%%%%%%%%%%%%%%%%%%%%%%%%%%%%%%%%%%%%%
% Subection 4.2 
% 
\subsection{How the knot coefficients of prime knots depends on the cylindrical radius}

\begin{figure}[htbp]
\begin{center}
  \includegraphics[width=0.9\hsize]{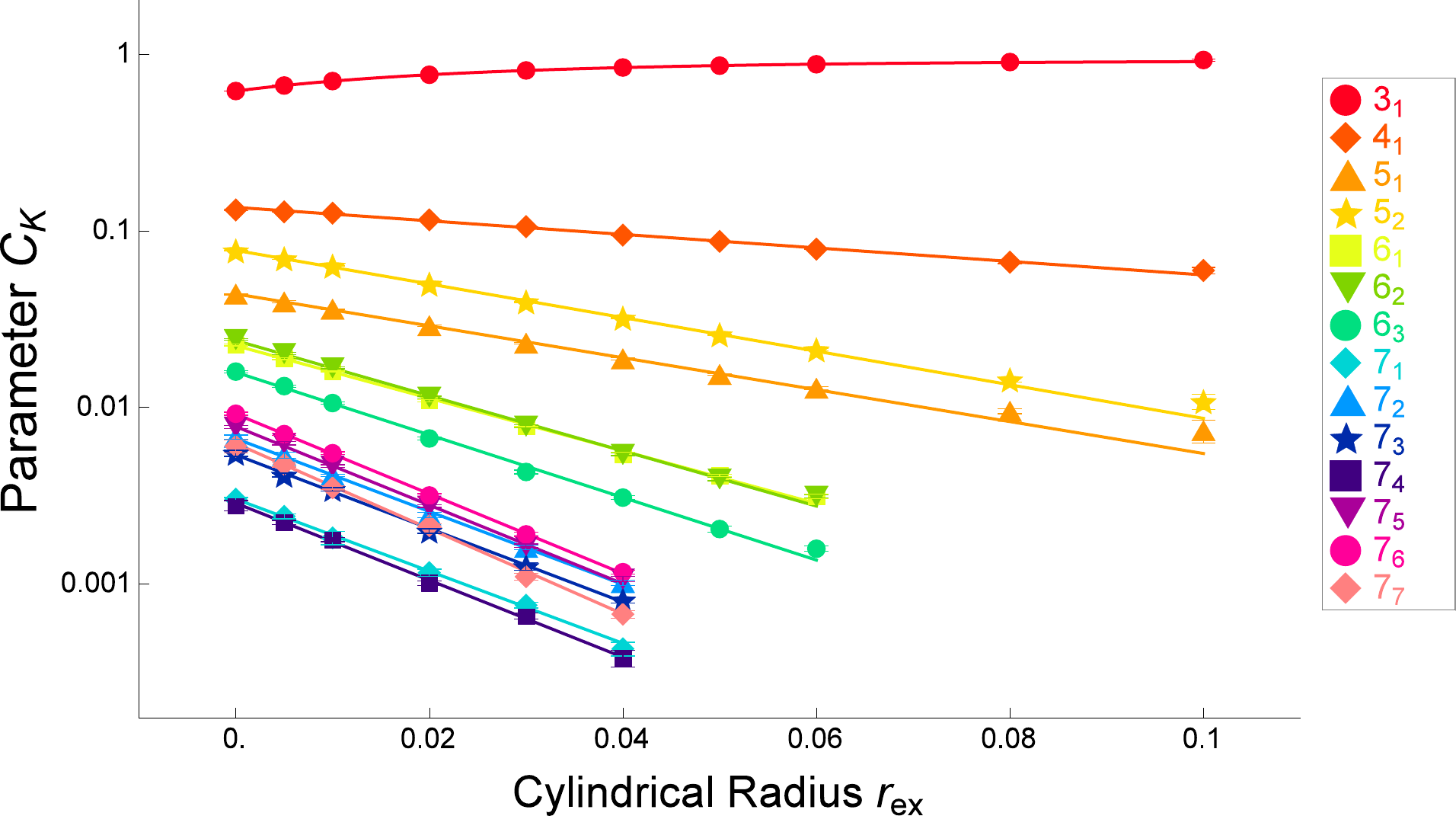}
\caption{Coefficients $C_K$ versus cylindrical radius $r_{\rm ex}$ for the primes knots with crossing number being less than or equal to seven in the semi-logarithmic scale. Fitted curves are given by applying Eq.  (\ref{eq:C31}) to the data of the trefoil knot and 
 (\ref{eq:CKprime}) to the data of the other prime knots. }
  \label{figCK}
\end{center}
\end{figure}

The coefficients $C_K$ of the prime knots $K$ with up to seven crossings are plotted against cylindrical radius $r_{\rm ex}$ in Fig. \ref{figCK} in the semi-logarithmic scale. 
%-----------------------------------------------------------------------
% Figure 6
%-----------------------------------------------------------------------
In Fig. \ref{figCK} it is clear that the coefficients $C_K$ for the prime knots other than the knot $3_1$ decay exponentially with respect to cylindrical radius $r_{\rm ex}$. Furthermore, as the minimal crossing number of a knot $K$ increases, the absolute value of the gradient of the fitted line to the data points increases.

Let us express the coefficient $C_K$ of a prime knot $K$ as a function of cylindrical radius $r_{\rm ex}$.  For any given prime knot other than the trefoil knot $3_1$ we introduce as a fitting formula an exponentially decaying function of cylindrical radius  $r_{\rm ex}$ as follows.  
%-----------------------------------------------------------------------
% Equation denoting the coefficient C_K
\begin{equation}
C_K(r_{\rm ex})=a_1(K) \exp(-b_1(K) r_{\rm ex})
\label{eq:CKprime}
\end{equation}
%-----------------------------------------------------------------------
Here, parameter $b_1(K)$ denotes the constant of exponential decay with respect to the cylindrical radius $r_{\rm ex}$, and  parameter $a_1(K)$  the coefficient for it. In Table \ref{Tab006} the best estimates of fitting parameters $a_1(K)$ and $b_1(K)$ of eq. (\ref{eq:CKprime}) together with the $\chi^2/{\rm DF}$ values are listed.
In Fig. \ref{figCK} the fitted curves are given by eq. (\ref{eq:CKprime}).
 We observe that the fitted curves are good since the $\chi^2$ values per DF are less than 2.0 
for all of them.

\begin{table}[htbp] 
\begin{center}
\begin{tabular}{c|ccc} 
\hline 
Knot type $K$ & $a_1(K)$ & $b_1(K) $ & $\chi^2/{\rm DF}$ \\ \hline
$4_1$ & $0.1357 \pm 0.0013$ & $8.82 \pm 0.27$ & $8.70$  \\ 
$5_1$ & $0.04387 \pm 0.00042$ & $20.81 \pm 0.34$ & $3.56$  \\ 
$5_2$ & $0.07741 \pm 0.00046$ & $21.91 \pm 0.22$ & $3.02$  \\ 
$6_1$ & $0.02234 \pm 0.00029$ & $34.30 \pm 0.56$ & $3.96$  \\ 
$6_2$ & $0.02389 \pm 0.00050$ & $35.97 \pm 0.82$ & $7.82$  \\ 
$6_3$ & $0.01580 \pm 0.00037$ & $40.8 \pm 1.1$ & $5.56$  \\ 
$7_1$ & $0.003022 \pm 0.000038$ & $47.02 \pm 0.79$ & $0.42$  \\ 
$7_2$ & $0.00665 \pm 0.00010$ & $47.52 \pm 0.82$ & $0.76$  \\ 
$7_3$ & $0.00538 \pm 0.00011$ & $48.0 \pm 1.2$ & $1.17$  \\ 
$7_4$ & $0.002866 \pm 0.000071$ & $50.2 \pm 1.5$ & $0.90$  \\ 
$7_5$ & $0.00778 \pm 0.00014$ & $51.22 \pm 0.99$ & $1.83$  \\ 
$7_6$ & $0.009159 \pm 0.000071$ & $52.02 \pm 0.44$ & $0.41$  \\ 
$7_7$ & $0.00624 \pm 0.00021$ & $55.6 \pm 1.7$ & $1.80$  \\ 
\hline
\end{tabular} 
\end{center}
\caption{Best estimates of parameters in eq. (\ref{eq:CKprime}) which expresses the coefficients $C_K$ of prime knots other than the trefoil knot as a function of cylindrical radius $r_{\rm ex}$. }
\label{Tab006}
\end{table}

In Table \ref{Tab006} we observe that as the minimal crossing number of knot $K$ increases the best estimate of $a_1$ of a knot $K$ becomes smaller while that of $b_1$ of a knot $K$ becomes larger. That is, if a knot $K$ is more complex than a given fixed knot, the knotting probability of the knot $K$ is smaller than that of the given fixed knot.

For the trefoil knot $3_1$ we express  
the coefficient of knotting probability, $C_{3_1}$, 
as a function of cylindrical radius $r_{\rm ex}$ by the following function: 
\begin{equation}
C_{3_1}(r_{\rm ex}) = a_0(3_1) (1 -a_1(3_1) \exp(- b_1(3_1) r_{\rm ex})) \, . 
\label{eq:C31}
\end{equation}
The coefficient of the trefoil knot $C_{3_1}$  approaches a constant value $a_0(3_1) \approx 0.92$ exponentially with respect to cylindrical radius $r_{\rm ex}$.  The best estimates are given by $a_0(3_1)= 0.919 \pm 0.003$, $a_1(3_1)=0.327 \pm 0.002$ and $b_1(3_1)= 33.1 \pm 0.8$ with $\chi^2/{\rm DF}= 1.8$. We conclude that the fitted curve is good since the $\chi^2$ values per DF is less than 2.0.

The coefficient $C_K$ of the trefoil knot $3_1$ increases gradually 
 from $C_{3_1}=0.62$ at $r_{\rm ex}=0$ to $C_{3_1}=0.91$ at $r_{\rm ex}=0.1$ 
as cylindrical  radius $r_{\rm ex}$ increases. The coefficient $C_{3_1}$ becomes relatively large as cylindrical radius $r_{\rm ex}$ becomes large.  
For an illustration, two examples of the ratios among coefficients $C_K$s are given as follows. 
%-----------------------------------------------------------------------
% Ratio of Parameter C_K 
\begin{equation}
C_{3_1}:C_{4_1}:C_{5_1}:C_{5_2} \sim 14: 3 : 1 : 1.8, \quad {\rm for} \,  r_{\rm ex}=0
\label{Eq007}
\end{equation}
%-----------------------------------------------------------------------
% Ratio of Parameter C_K
\begin{equation}
C_{31}:C_{41}:C_{51}:C_{52}\sim119:7:1: 1.4,  \quad {\rm for} \,  r_{\rm ex}=0.1
\label{Eq008}
\end{equation}
%-----------------------------------------------------------------------
Thus, the maximum probability of knot $3_1$ is $119$ times larger  than that of knot $5_1$ in the case of  $r_{\rm ex}=0.1$.

We thus suggest that the maximum value of the knotting probability of knot $3_1$ does not decrease even if the excluded volume becomes very large. Here we recall that the maximum knotting probability of a prime knot $K$ is almost determined by the coefficient $C_K$, and that coefficient of knot $3_1$ increases gradually as the cylindrical radius $r_{\rm ex}$ increases, and hence a large number of SAPs being equivalent to 
knot $3_1$ or its composite knots  are generated when the cylindrical radius $r_{\rm ex}$ is large.  
When the cylindrical radius is large such as $r_{\rm ex}=0.1$, 
the majority of nontrivial knots are given by the trefoil knot and its composite knots.

%%%%%%%%%%%%%%%%%%%%%%%%%%%%%%%%%%%%%%%%%%%%%%%%%%%%%%%
% subsection 4.3 
%
\subsection{Factorization of knot coefficients for composite knots }

We now show numerically that the factorization property of knot coefficients $C_K$ holds  
for such composite knots $K$ that consist of two prime knots ($n=2$) and  
three prime knots ($n=3$), respectively.  We shall show it 
for the best estimates of the knot coefficients $C_K$ of the fitted curves 
given by Eq. (\ref{eq:4formula}) applied to the data of the knotting probabilities.

Let us recall the  factorization property of knot coefficients $C_K$ explicitly in the case of $n=2$:  If a knot $K$ is given by the product of two different prime knots $K_1$ and $K_2$, the coefficient of the composite knot $K=K_1 \# K_2$ is given by 
\begin{equation}
C_{K_1 \# K_2} = C_{K_1} C_{K_2} ,   \label{eq:fac1}
\end{equation}
while if it consists of the product of a pair of the same prime knot $K_1$,  the coefficient of the composite knot $K_1 \# K_1$ is given by 
\begin{equation}
C_{K_1 \# K_1} = {C_{K_1}^2}/{2 !}  \, . \label{eq:fac2}
\end{equation}

Assuming the factorization properties (\ref{eq:fac1}) and (\ref{eq:fac2})  for the coefficients $C_K$ of composite knots $K=K_1 \# K_2$ consisting of two prime knots $K_1$ and $K_2$, we can  evaluate numerically the coefficients $C_K$ of the  composite knots $K$. Here we make use of expressions (\ref{eq:CKprime}) and (\ref{eq:C31}) as functions of cylindrical radius $r_{\rm ex}$, 
and taking their products.  For instance, in the case of the composite knot of two figure-eight knots $K=4_1 \# 4_1$ we evaluate the coefficient $C_{4_1 \# 4_1}$ by 
\begin{equation} 
 C_{4_1 \# 4_1}(r_{\rm ex})  = c_1(4_1)^2 \exp\left( - 2 d_1(4_1) r_{\rm ex} \right)/2! . 
\label{eq:ab4141}
\end{equation}
 
We thus have two methods for evaluating knot coefficients $C_K$ for such composite knots 
$K$ consisting of two prime knots $K_1$ and $K_2$.  In the first method, by applying Eq. (\ref{eq:4formula}) we derive fitted curves to the data points of the knotting probabilities of composite knots $K=K_1 \# K_2$, and evaluate coefficients $C_K$ by the best estimates of the fitting parameters of Eq. (\ref{eq:4formula}).  In the second method, we evaluate knot coefficients $C_K$ for composite knots consisting of prime knots $K_1$ and $K_2$  
by taking the product of the coefficients $C_{K_1}$ and  $C_{K_2}$ 
given by (\ref{eq:CKprime}) and (\ref{eq:C31}) as functions of cylindrical radius.

% -----------------------------
% Figure 7. 
%
\begin{figure}[htbp]
\begin{center}
  \includegraphics[width=1.0\hsize]{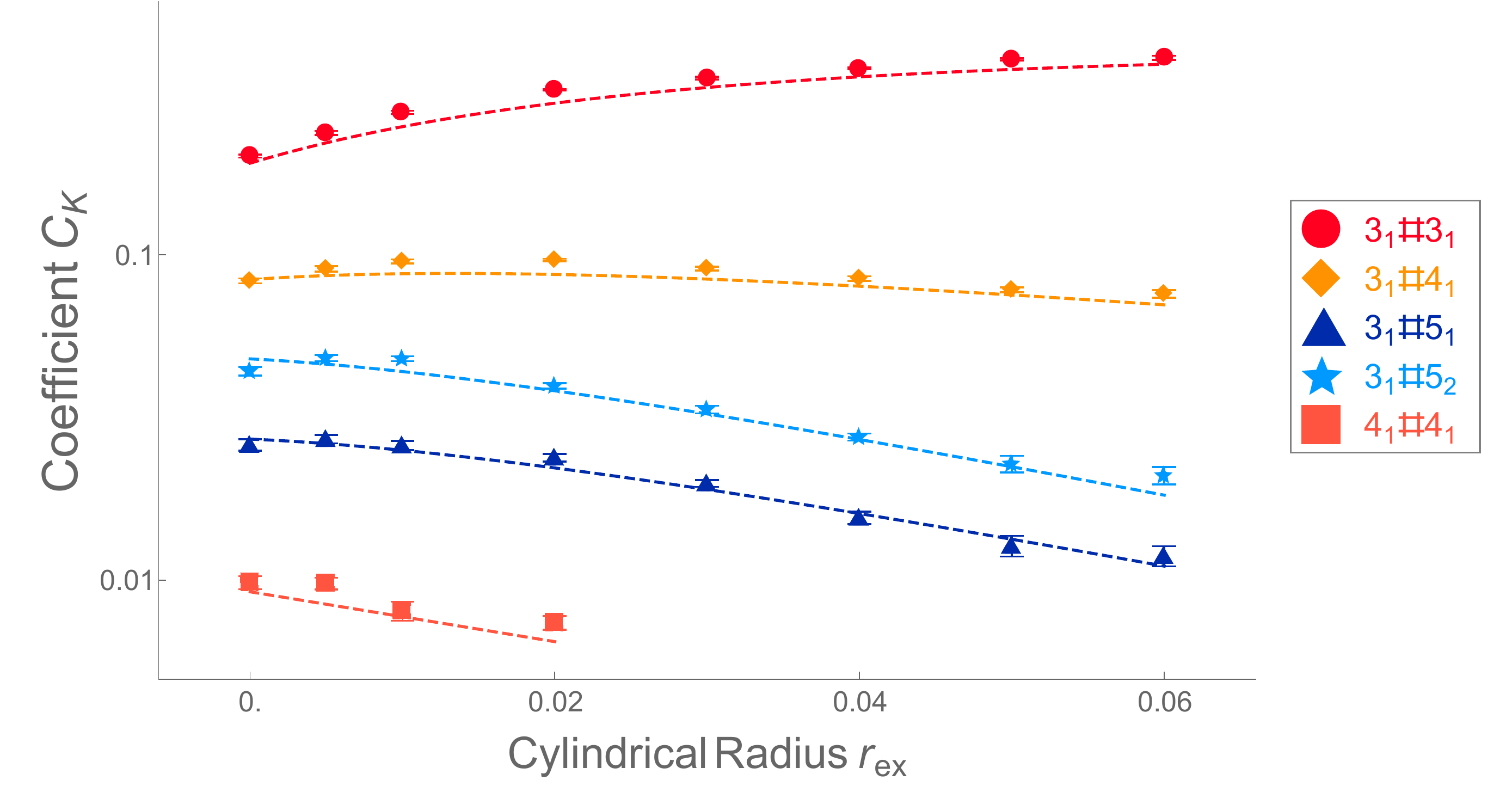}
\caption{Coefficient $C_K$ of a composite knot $K=K_1 \# K_2$ versus cylindrical radius $r_{\rm ex}$. 
The data points for knots $3_1 \# 3_1$,  $3_1 \# 4_1$, $3_1 \# 5_1$, $3_1 \# 5_2$,  and $4_1 \# 4_1$ are depicted by filled circles, diamonds, upper triangles, stars, and squares, respectively. 
They are obtained from the fitted curves given by Eq. (\ref{eq:4formula}).  
To  each composite knot $K=K_1 \# K_2$  the dotted curve is given by  the product  of $C_{K_1}$ and $C_{K_2}$ as functions of cylindrical radius $r_{\rm ex}$ given in Eqs. (\ref{eq:CKprime}) 
 and (\ref{eq:C31}).   }
  \label{figCKCK}
\end{center}
\end{figure}
%-----------------------------------------------------------------------

In Fig. \ref{figCKCK} we plot the estimates of the coefficient $C_K$ of a composite knot $K=K_1 \# K_2$ consisting of two prime knots $K_1$ and $ K_2$ against cylindrical radius $r_{\rm ex}$. 
They are shown by the data points in Fig. \ref{figCKCK}. Here we recall that the best estimate of $C_K$ is evaluated by applying Eq. (\ref{eq:4formula}) to the knotting probabilities of the composite knot $K$ plotted against segment number $N$ 
for the cylindrical SAP with a given value of radius $r_{\rm ex}$.
To each of the five composite knots, the dotted curve is drawn by making use of the expression  as a function of cylindrical radius $r_{\rm ex}$ given by the product of Eqs. (\ref{eq:CKprime}) and (\ref{eq:C31}). For instance, we recall (\ref{eq:ab4141}) for the composite knot of two figure-eight knots. 

The dotted curves are very close to the data points and almost overlapping them, 
as shown in In Fig. \ref{figCKCK}. 
The agreement of the results of the two method should be quite remarkable. 
Here we remark that they have no parameters to fit. 

We have thus numerically confirmed that the factorization properties (\ref{eq:fac1}) and  (\ref{eq:fac2}) for $n=2$ hold among the best estimates of  coefficients $C_K$ in the cylindrical SAP with various different values of cylindrical radius.

% -----------------------------------------
% Figure 8.
%
\begin{figure}[htbp]
\begin{center}
\includegraphics[width=1.0\hsize]{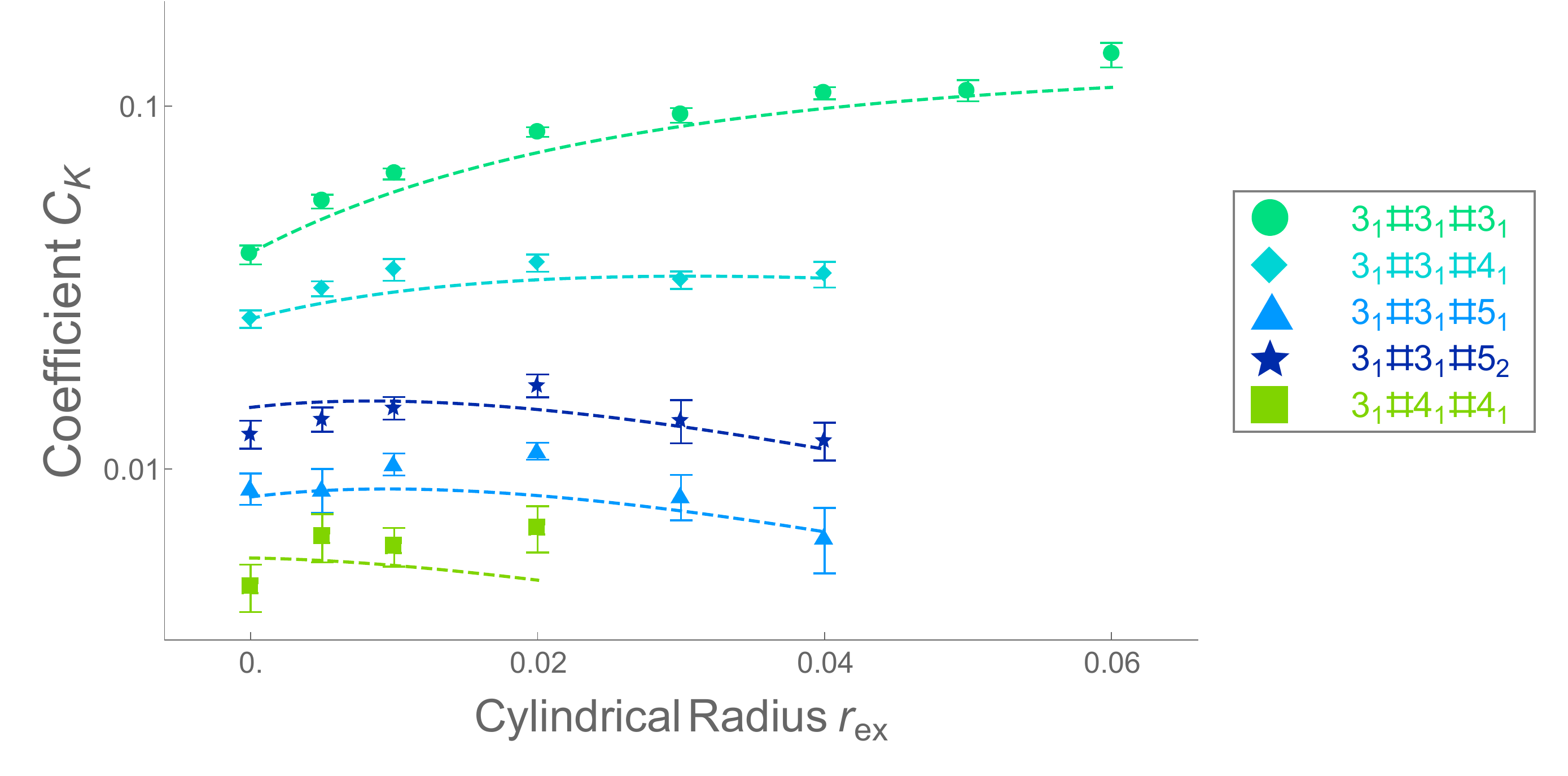}
 \caption{Coefficients $C_K$ for composite knots $K$ consisting of three prime knots: $K_1 \# K_2 \# K_3$ versus cylindrical radius $r_{\rm ex}$. The data points for knots $3_1 \# 3_1 \# 3_1$, $3_1 \# 3_1 \# 4_1$, $3_1 \# 3_1 \# 5_1$, $3_1 \# 3_1 \# 5_2$, and $3_1 \# 4_1 \# 4_1$ are depicted by filled circles, diamonds, upper triangles, stars, and squares, respectively. 
They are evaluated by applying Eq. (\ref{eq:4formula}) to the data of the knotting probabilities.  
The dotted curves are given by taking the product of the coefficients $C_K$ of the constituent prime knots. For instance, the coefficient $C_{3_1 \# 3_1 \# 3_1}$ is calculated as $\left( C_{3_1}\right)^3/3!$.           
}
  \label{figCKCKCK}
\end{center}
\end{figure}
%-----------------------------------------------------------------------

Also for such composite knots $K$ that consist of three prime knots 
as $K=K_1 \# K_2 \# K_3$,  we have the same two methods for evaluating the knot coefficients $C_K$.   
First, by applying Eq. (\ref{eq:4formula}) we derive fitted curves to the data points of the knotting probabilities of composite knots $K=K_1 \# K_2 \# K_3$ plotted against segment number $N$,  
we evaluate coefficients $C_K$ as the best estimates of the  fitting parameter. 
Second, we evaluate $C_K$ by taking the product 
of $C_{K_1}$, $C_{K_2}$ and  $C_{K_3}$ given by Eqs. (\ref{eq:CKprime}) and (\ref{eq:C31}).

In Fig. \ref{figCKCKCK} 
 we plot the estimates of the coefficient $C_K$ of a composite knot $K=K_1 \# K_2 \# K_3$ consisting of three prime knots $K_1$, $ K_2$ and $K_3$ against cylindrical radius $r_{\rm ex}$. Here we recall that the best estimate of $C_K$ for a composite knot $K$ obtained by applying Eq. (\ref{eq:4formula})  to the knotting probability of the composite knot $K$ for the cylindrical SAP with a given value of radius $r_{\rm ex}$, and it  is shown by a data point in Fig. \ref{figCKCKCK}.  
To each of the composite knots $K$
the dotted curve is drawn by making use of the expressions as functions of cylindrical radius $r_{\rm ex}$ given by Eqs. (\ref{eq:CKprime}) and (\ref{eq:C31}).  The dotted curves are drawn by calculating the numerical values against radius $r_{\rm ex}$  by making use of the coefficients of the prime knots as functions of cylindrical radius $r_{\rm ex}$ and do not have any fitting parameters in order to make them fit to the data points of Fig. \ref{figCKCKCK}. However, the dotted curves are very close to the data points.

We have thus numerically  confirmed that the factorization properties 
(\ref{eq:fact3}) for $n=3$ hold in the cylindrical SAP 
for various values of cylindrical radius. 
 
%%%%%%%%%%%%%%%%%%%%%%%%
%
%  Sec 5 
%
\section{Various properties of the fitting parameters for knotting probability}
%%%%%%%%%%%%%%%%%%%%%%%%%%%%%%%%%%%%%%%%%
\subsection{Thickness dependence of the knotting probability in the small 
segment number region}  

We show the connection of the thickness dependence of knot coefficients 
$C_K$ to the gradient of the knotting probability as a linear function of 
segment number $N$ in the small-$N$ region studied in Refs. \onlinecite{Rybenkov,Shaw-Wang}. 

If the number of segments $N$ is much smaller than the characteristic length of knotting probability $N_0$, then the four-parameter formula is approximated by a linear function of $N$ as Eq.  (\ref{eq:small-N}). Here, the gradient of the graph as a function of $N$ is given by $C_K/N_K$ and it is approximately expressed as a function of cylindrical radius $r_{\rm ex}$ as 
\begin{equation}
{\frac {C_{3_1} (r_{\rm ex} )}  { N_{3_1}(r_{\rm ex}) } }
= \frac { a_0(3_1) (1-   a_1(3_1)  \exp(- b_1(3_1) r_{\rm ex})  ) } 
            {c_0(3_1) + c_1(3_1) \exp(d_1(3_1) r_{\rm ex}) }  
\label{eq:grad31}
\end{equation} 
for the trefoil knot and 
\begin{equation}
{\frac {C_K(r_{\rm ex} )}  {N_K(r_{\rm ex} ) }}
= \frac {a_1(K) \exp(- b_1(K) r_{\rm ex})} {c_0(K) + c_1(K) \exp(d_1(K) r_{\rm ex})}  
\label{eq:gradient}
\end{equation} 
for prime knots other than the trefoil knot. 
 
In Refs. \cite{Rybenkov,Shaw-Wang} the graph of the knotting probability of a knot against segment number $N$ is expressed as a linear function of $N$ and its gradient  is approximated by an exponentially decaying function of the diameter of cylindrical segments.  Here the gradient corresponds to the ratio 
$C_K/N_K$ in the present paper. The thickness dependence of the gradient $C_K/N_K$ expressed  as eq. (\ref{eq:gradient}) is more complex than an exponentially decaying function.

We can show that the expressions of Eq. (\ref{eq:grad31}) and (\ref{eq:gradient}) for the gradient $C_K/N_K$ generalize the thickness dependence given in Refs. \cite{Rybenkov,Shaw-Wang} valid in  the small-$N$ region to that of  much wider ranges of segment number $N$:   The expressions (\ref{eq:grad31}) and (\ref{eq:gradient}) coincide with the approximate exponential dependence of the knotting probability on the thickness of cylindrical segments shown in Refs. \cite{Rybenkov,Shaw-Wang} for small $N$,  and they are valid also for large $N$. 

Let us express the approximate exponential-dependence  as $\exp( - \gamma_K r_{\rm ex})$  in terms of cylindrical radius $r_{\rm ex}$. In Ref. \cite{Rybenkov} the decay constants $\gamma_K$ are estimated by 44, 62, 84 and 84 for the trefoil knot $3_1$, the figure-eight knot $4_1$, knot $5_1$ and knot $5_2$, respectively.  
It is easy to show explicitly that the graph of  $C_K(r_{\rm ex} ) / N_K (r_{\rm ex} )$ versus cylindrical radius $r_{\rm ex}$ almost completely overlaps with that of  $ \left( C_K(0) / N_K (0) \right) \exp( - \gamma_K r_{\rm ex}) $ versus  cylindrical radius $r_{\rm ex}$ for each knot $K$ among the four prime knots: $3_1$, $4_1$, $5_1$ and $5_2$.

%***********************************************************************
% Subsection: 5.2 
%  
%***********************************************************************
\subsection{Connection to lattice knots: Universal ratios of knotting probabilities 
of prime knots }

It is numerically shown in Ref. \cite{Rechnitzer} that the asymptotic behavior of the ratio of knotting probabilities does not depend on the types of lattices for some knots.  It is suggested that the ratio of the knotting probability of a knot $K_1$  to that of a knot $K_2$ is given by  a universal value for each pair of knots $K_1$ and $K_2$. From the viewpoint of formula (\ref{eq:4formula}) the knotting probability ratio in the large $N$ limit is expressed in terms of the ratio of coefficients $C_K$: 
\begin{eqnarray}
{\frac {P(N, r_{\rm ex}, K_1)} {P(N, r_{\rm ex}, K_2)}} = 
\frac {C_{K_1}} {C_{K_2}} . 
\end{eqnarray}

Let us now make an application of Eq. (\ref{eq:CKprime}) for 
expressing the thickness dependence of knot coefficents $C_K(r_{\rm ex})$. 
Here we recall that the best estimates of are given in Table \ref{Tab006}. The knotting probability ratio of $K={4_1}$ to $5_1$ is given by 15 in Ref. \cite{Rechnitzer}. Applying formula  (\ref{eq:CKprime}) we have   
\begin{equation}
\frac {C_{4_1}} {C_{5_1}} = \frac {0.136 \exp(-8.8 \, r_{\rm ex})} 
{0.044 \exp( - 21.0 \, r_{\rm ex})}  = 3.09  \exp(12.2 \, r_{\rm ex})  \, .  
\end{equation}
By equating it to the ratio 15 we have the estimate of the corresponding cylindrical radius 
\begin{equation}
r_{\rm ex} = 0.13 .   
\end{equation} 
Similarly, for knot probability ratio of $K=4_1$ to $5_2$, 
by setting the equation: $C_{4_1}/C_{5_2}= 9$ we have 
\begin{equation}
r_{\rm ex} = 0.12 .  
\end{equation}
It is interesting to note that both of the values of radius $r_{\rm ex}$ are close to 1/8. 
Here, the diameter $2 r_{\rm ex}$ is given by 1/4. 

We suggest that the results of the cylindrical SAP with $r_{\rm ex}=1/8$ are consistent with those of lattice SAP.  We have a conjecture that if we evaluate the statistical length of SAP on lattice, then the characteristic length of knotting probability for lattice SAP corresponds to that of off-lattice SAP with $r_{\rm ex}=1/8$.

% ---------------------------------------------------------------
% subsection 5.3  
%
\subsection{Sum rules of knot coefficients and factorization properties}

We now argue that knot coefficients $C_K$ satisfy simple relations, which we call sum rules. Furthermore, we show that the sum rules are consistent with the factorization properties of coefficients $C_K$ for composite knots.

Let us classify all knots into classes of knots consisting of $n$ prime knots. 
For instance, we have $n=1$ for prime knots, $n=2$ for composite knots consisting of two prime knots, etc.  Suppose that a given composite knot $K$ consists of $n$ prime knots. Then, we denote the number of constituent prime knots $n$ by $|K|$, i.e. $|K|=n$. For simplicity, we assume that the characteristic lengths $N_K$ for knots $K$ are given by the same number $N_0$, the exponent $m(K)$ is given by the number of such  prime knots that are constituent knots of $K$, the finite-size corrections $\Delta N(K)$ are  given by zero. 

We remark that the sum of the knotting probabilities over all knots is given by 1.  We therefore have 
\begin{eqnarray}
1 & = & \sum_{n=0}^{\infty} \sum_{|K|=n} P_K(N)  \nonumber \\ 
& = & \sum_{n=0}^{\infty} \sum_{|K|=n} C_K x^{n} \exp(-x) \, . \label{eq:sumPK}  
\end{eqnarray}
Here variable $x$ is defined by $x=N/N_0$. The symbol $\sum_{|K|=n}$ denotes the sum over all such composite knots that consist of $n$ prime knots. It follows from Eq. (\ref{eq:sumPK}) that we have 
\begin{equation}
e^{x} = \sum_{n=0}^{\infty}  x^{n} \sum_{|K|=n} C_K \, . 
\end{equation} 
Through the Taylor expansion of the exponential function  
we have an infinite number of conditions for the knot coefficients $C_K$  
\begin{equation}
 \sum_{|K|=n} C_K = 1/n ! .  \label{eq:sum-rule}  
\end{equation}
The sum of  the knot coefficients over such knots that consist of $n$ prime knots 
is given by $1/n!$. 

In the case of $n=2$, by expressing the composite knot $B$ as $K=K_1 \# K_2$ we have 
\begin{eqnarray} 
& & \sum_{|K|=2} C_K  =  \sum_{K_1<K_2; prime} C_{K_1 \# K_2} +  \sum_{K_1 =K_2: prime} C_{K_1 \# K_2} 
\nonumber \\ 
 & = & {\frac 1 2}  \sum_{K_1: prime} \sum_{K_2 \ne K_1: prime} C_{K_1 \# K_2} 
+  \sum_{K_1 =K_2: prime} C_{K_1 \# K_2} \nonumber \\
\end{eqnarray} 
If we assume for different constituent knots $K_1 \ne K_2$ we have 
%\begin{equation}
$ C_{K_1 \# K_2} =  C_{K_1} C_{K_2}$ and  for the same knot we have
%\quad  
$   C_{K_1 \# K_1} =  C_{K_1}^2 /2! $,    
%\end{equation}
then we have 
\begin{equation}
\sum_{|K|=2} C_K = \left( \sum_{K_1: prime} C_{K_1} \right)^2 / 2! 
\end{equation} 
Thus, the sum rule for $n=2$ (i.e., the condition for $n=2$) is derived  
if the sum of the coefficients $C_{K_1}$ of prime knots $K_1$  
over all the prime knots is given by 1:  
\begin{equation} 
\sum_{K_1: prime} C_{K_1}  = 1 .  \label{eq:sum-prime}
\end{equation}
We call Eq.  (\ref{eq:sum-prime}) the sum rule for $n=1$.  

If a composite knot $K$ consists of  $n$ prime knots where $n= n_1 + n_2 + \cdots $ and $n_1$ knots are given by the same prime knot $K_1$, $n_2$ knots by $K_2$, etc., we have factorization properties (\ref{eq:fact3}) as follows.  
\begin{equation}
C_K = \left( C_{K_1} \right) ^{n_1}/n_1 !  \, \cdot \,  \left( C_{K_2} \right) ^{n_2}/n_2 !\cdots . 
\label{eq:facn}
\end{equation}
Then,  the sum of coefficients $C_K$ over all composite knots 
consisting of $n$ prime knots is expressed with  the sum of coefficients $C_K$ over all prime knots as 
\begin{equation}
\sum_{|K|=n} C_K = \left( \sum_{K_1: prime} C_{K_1} \right)^n / n ! 
\end{equation} 
It follows that all the conditions (\ref{eq:sum-rule}) are derived from the condition (\ref{eq:sum-prime}) that the sum of coefficients $C_{K_1}$ over all prime knots $K_1$ is given by 1.

% ------------------------------------------
%Table 5
%
\begin{table}[htbp]
\begin{center} 
\begin{tabular}{c|c}
\hline 
$r_{\rm ex}$ & Sum of prime knot coefficients: $\sum_{|K|=1} C_K$ 
%over some prime knots 
\\
\hline 
\hline 
0.0  & 
$ 0.9737 \pm 0.0049$
%$ 0.97375307 \pm 0.00492$
%(upto seven crossings)  
\\ 
0.005 & 
$0.9855  \pm 0.0050$ 
%$0.98551783  \pm 0.00498$ 
%(upto seven crossings) 
\\ 
0.01 & 
$0.9967 \pm 0.0038$
%$0.99674881 \pm 0.003768$
%(upto seven crossings) 
\\ 
0.02 & 
$1.0021 \pm 0.0027$ 
%$1.00215139 \pm 0.002728$ 
%(upto seven crossings) 
\\ 
0.03 & 
$1.0056  \pm 0.0029$
%$1.005664156  \pm 0.002879$
%(upto seven crossings) 
\\ 
0.04 & 
$1.0051 \pm 0.0029$ 
%$1.005100225 \pm 0.00285$ 
%(upto seven crossings) 
\\ 
0.05 & 
$0.9990 \pm 0.0022$
%$0.99906959 \pm 0.002216$
%(upto six crossings) 
\\ 
0.06 & 
$0.9969 \pm 0.0024$
%$0.99697509 \pm 0.002419$
%(upto six crossings) 
\\ 
0.08 & 
$0.9921 \pm 0.0049$ 
%$0.99212725 \pm 0.00493$ 
%(upto five crossings) 
\\ 
0.1 & 
$1.0039 \pm 0.0166$ 
%$1.00395265 \pm 0.0166$ 
%(upto five crossings) 
\\ 
\hline 
\end{tabular}
\end{center}
\caption{The sum of coefficients $C_K$ over some prime knots  
 for cylindrical SAPs with ten different values of cylindrical radius $r_{\rm ex}$. We take the sum over the  prime knots with up to seven  crossings from $r_{\rm ex}=0$ to $r_{\rm ex}=0.04$, up to six crossings for $r_{\rm ex}=0.05, 0.06$,  and up to five crossings for $r_{\rm ex}=0.08, 0.10$.  
} \label{tab:sumrule}
\end{table}

Let us confirm the sume rule (\ref{eq:sum-prime}) numerically. In Table \ref{tab:sumrule} we present the sum of the best estimates of coefficients $C_K$ over the prime knots with up to some number of crossings for the cylindrical SAP with radius $r_{\rm ex}$ in each value of  radius among the ten different values. 
We observe that the sum rule (\ref{eq:sum-prime}) is numerically satisfied with respect to errors by the coefficients $C_K$ of the prime knots we have investigated.

It is impressive to see how well the sum rule holds with respect to errors, although we do not assume that $m(K)=1$ or $\Delta N(K)= 0$ for all prime knots.  
Here we recall that the sum rule is derived from the knotting probability formula (\ref{eq:4formula}) 
by assuming some properties of the fitting parameters such as the factorization properties of exponents $m(K)$ and coefficients $C_K$.

We suggest that the confirmation of the sum rule for prime knots (i.e., $n=1$) 
shown in Table \ref{tab:sumrule}  gives a  numerical  support for the validity of the local knot picture in the knotting probability.  
Here we assume that the factorization properties of knot exponents and knot coefficients 
for the knotting probability are derived from the local knot picture 
that the knotted region of a knotted SAP is localized in the whole configuration of the SAP.  

%%%%%%%%%%%%%%%%%%%%%%%%%%%%%%%%%%%
%
% Sec 6 
%
\section{Asymptotic expansion of the knotting probability}

We now investigate how far the three-parameter asymptotic formula (\ref{eq:3formula}) 
is effective for describing the knotting probability as a function of the segment number $N$. Here we recall that the four-parameter fitting formula (\ref{eq:4formula}) is derived by modifying the asymptotic formula (\ref{eq:3formula}) and also that it corresponds to the asymptotic expansion (\ref{eq:ZK}) of the logarithm of the partition function $Z_K(N)$ of a RP or SAP with  fixed knot $K$ with respect to the inverse of the segment number  $N$. We also recall that the ratio of the partition function $Z_K(N)$ to that of no topological constraint $Z(N)$ corresponds  to the knotting probability $P_K(N)$ of the knot $K$.

%-----------------------------------------------------------------------
% Table 3. Best estimates of knotting probabilities of Nontrivial knot with three parameters
%
\begin{table}[htbp]
\begin{center}
\begin{tabular}{c|cccc}
\hline 
$r_{\rm ex}$ & $C_K$ & $m(K)$ & $N_K$ & $\chi^2/$DF \\ \hline 
\multicolumn{5}{c}{Knot $3_1$} \\ \hline 
 0 & $0.6201 \pm 0.0018$ & $0.9649 \pm 0.0094$ & $248.9 \pm 1.3$ & $2.21$  \\
 0.005 & $0.6667 \pm 0.0027$ & $0.949 \pm 0.011$ & $365.5 \pm 2.4$ & $4.31$  \\
 0.01 & $0.707 \pm 0.0024$ & $0.9455 \pm 0.0086$ & $499.4 \pm 3.0$ & $3.94$  \\
 0.02 & $0.7671 \pm 0.0017$ & $0.9428 \pm 0.0053$ & $847.2 \pm 3.8$ & $2.52$  \\
 0.03 & $0.8108 \pm 0.0013$ & $0.9575 \pm 0.0041$ & $1327.1 \pm 5.5$ & $1.80$  \\
 0.04 & $0.8422 \pm 0.0012$ & $0.9576 \pm 0.0033$ & $2023.7 \pm 7.9$ & $1.43$  \\
 0.05 & $0.8654 \pm 0.0016$ & $0.967 \pm 0.0041$ & $2973. \pm 17.$ & $2.24$  \\
 0.06 & $0.8819 \pm 0.0016$ & $0.9665 \pm 0.0035$ & $4328. \pm 26.$ & $1.57$  \\
\hline
%\end{tabular} \end{table}
%\begin{table}[htbp] \begin{tabular}{c|cccc}
\multicolumn{5}{c}{Knot $4_1$} \\ \hline 
%$r_{\rm ex}$ & $C_K$ & $m(K)$ & $N_K$ & $\chi^2/$DF \\ \hline 
0 & $0.13072 \pm 0.00091$ & $1.05 \pm 0.024$ & $242.4 \pm 3.2$ & $2.99$  \\
0.005 & $0.12851 \pm 0.001$ & $1.01 \pm 0.021$ & $355.4 \pm 4.6$ & $3.27$  \\
0.01 & $0.12529 \pm 0.00069$ & $1.029 \pm 0.015$ & $475.1 \pm 4.9$ & $2.11$  \\
0.02 & $0.11585 \pm 0.00057$ & $0.99 \pm 0.013$ & $818.8 \pm 8.6$ & $2.16$  \\
0.03 & $0.10636 \pm 0.00054$ & $1.003 \pm 0.014$ & $1275. \pm 17.$ & $2.54$  \\
0.04 & $0.09585 \pm 0.0005$ & $0.982 \pm 0.013$ & $1964. \pm 29.$ & $2.33$  \\
0.05 & $0.08798 \pm 0.00037$ & $0.9979 \pm 0.0095$ & $2852. \pm 37.$ & $1.20$  \\
0.06 & $0.07966 \pm 0.00049$ & $0.989 \pm 0.012$ & $4166. \pm 85.$ & $1.70$  \\
\hline 
\end{tabular}
\end{center}
\caption{Best estimates of  three-parameter formula (\ref{eq:3formula})
to the knotting probability of the trefoil knot $3_1$ and the figure-eight knot $4_1$ 
for cylindrical SAP with radius $r_{\rm ex}$. }
\label{tab:3para3141}
\end{table}

\begin{table}[htbp]
\begin{center}
\begin{tabular}{c|cccc}
\hline 
$r_{\rm ex}$ & $C_K$ & $m(K)$ & $N_K$ & $\chi^2/$DF \\ \hline 
\multicolumn{5}{c}{Knot $5_1$} \\ \hline 
0 & $0.04261 \pm 0.00041$ & $1.181 \pm 0.034$ & $235.3 \pm 4.2$ & $1.95$  \\
0.005 & $0.03942 \pm 0.00053$ & $1.076 \pm 0.039$ & $350.1 \pm 8.1$ & $3.18$  \\
0.01 & $0.03565 \pm 0.00025$ & $1.072 \pm 0.02$ & $473.4 \pm 6.4$ & $1.03$  \\
0.02 & $0.02899 \pm 0.00025$ & $1.053 \pm 0.024$ & $797. \pm 15.$ & $1.88$  \\
0.03 & $0.02337 \pm 0.0002$ & $1.04 \pm 0.024$ & $1237. \pm 28.$ & $1.76$  \\
0.04 & $0.01897 \pm 0.0002$ & $1.041 \pm 0.028$ & $1857. \pm 56.$ & $2.16$  \\
0.05 & $0.01562 \pm 0.00015$ & $1.057 \pm 0.023$ & $2675. \pm 79.$ & $1.22$  \\
0.06 & $0.01298 \pm 0.0002$ & $1.058 \pm 0.033$ & $3840. \pm 190.$ & $1.92$  \\
\hline 
%\end{tabular} \caption{$5_1$ knot} \end{table}
%\begin{table}[htbp] \begin{tabular}{c|cccc}
\multicolumn{5}{c}{Knot $5_2$} \\ \hline 
%$r_{\rm ex}$ & $C_K$ & $m(K)$ & $N_K$ & $\chi^2/$DF \\ \hline 
0 & $0.07434 \pm 0.00067$ & $1.214 \pm 0.031$ & $228.4 \pm 3.6$ & $2.76$  \\
0.005 & $0.06843 \pm 0.00061$ & $1.106 \pm 0.026$ & $343.0 \pm 5.2$ & $2.53$  \\
0.01 & $0.06229 \pm 0.00044$ & $1.092 \pm 0.021$ & $461.3 \pm 6.2$ & $1.84$  \\
0.02 & $0.04995 \pm 0.00035$ & $1.045 \pm 0.019$ & $792. \pm 12.$ & $2.03$  \\
0.03 & $0.03991 \pm 0.00031$ & $1.060 \pm 0.023$ & $1234. \pm 27.$ & $2.61$  \\
0.04 & $0.0322 \pm 0.0002$ & $1.046 \pm 0.017$ & $1848. \pm 34.$ & $1.36$  \\
0.05 & $0.0262 \pm 0.00021$ & $1.049 \pm 0.020$ & $2744. \pm 71.$ & $1.53$  \\
0.06 & $0.02168 \pm 0.00028$ & $1.061 \pm 0.028$ & $3850. \pm 160.$ & $2.24$  \\
\hline 
\end{tabular}
\end{center}
\caption{Best estimates of three-parameter formula (\ref{eq:3formula})
to the knotting probabilities of  knots $5_1$ and $5_2$ 
for cylindrical SAP with radius $r_{\rm ex}$. }
\label{tab:3para5152}
\end{table}

It seems that the results of the asymptotic formula are closer to those of  on-lattice models of SAP than those of the four-parameter formula (\ref{eq:4formula}). 
We apply the asymptotic formula (\ref{eq:3formula}) to the data of the knotting probabilities 
of various knots plotted against  the number of segments $N$ for the cylindrical SAP with several different values of cylindrical radius $r_{\rm ex}$. The best estimates of the parameters of Eq.  (\ref{eq:3formula}) are listed in Tables \ref{tab:3para3141} and \ref{tab:3para5152} in Appendix B for the four knots: $3_1$, $4_1$, $5_1$ and $5_2$. 
We find that the $\chi^2$ values of the fitted curves are larger than those of the four-parameter formula (\ref{eq:4formula}) but they are not very large.  Furthermore, the estimates of the exponent $m(K)$ for prime knots are closer to 1 than those of the four-parameter formula (\ref{eq:4formula}). 
Here we recall that 
the estimates of the exponent $m(K)$ are given by integers for on-lattice SAP.

We suggest that the finite-size effect is significant in the knotting probability for off-lattice models of RP or SAP. If we neglect the small-$N$ region and consider only the large-$N$ region we expect that the knotting probability 
is well approximated by the  asymptotic formula  (\ref{eq:3formula}). 
However, for the off-lattice models with the number of segments from $N=100$ to $N=1000$, the finite-size effects are important, and hence the knotting probability is well approximated by  the four-parameter formula (\ref{eq:4formula}) in which we take into account corrections due to the finite-size effect.

%***********************************************************************
% Section: 7. 
% Conclusion
%***********************************************************************
\section{Concluding Remarks}

The knotting probability is significant in the topological properties of knotted ring polymers in solution. In particular, the characteristic length of the knotting probability plays a central role not only in the knotting probability but also in the scaling behavior of the RP or SAP under a topological constraint.

We have studied the dependence of the knotting probability on the thickness of cylindrical segments for the cylindrical SAP. We have numerically shown that the maximum value of the knotting probability of the trefoil knot  increases with respect to the radius of cylindrical segments, $r_{\rm ex}$, while those of other prime knots  decrease exponentially with respect to it. 
Here we recall that cylindrical radius $r_{\rm ex}$ is  the excluded-volume parameter of the cylindrical SAP. 

We expect that the results of the knotting probability for the cylindrical SAP are useful for studying the knotting probabilities for various other models of semi-flexible ring polymers. 
Through results of the cylindrical SAP we can predict possible topological effects in other models of RP or SAP. For instance, the cylindrical SAP model interpolates off-lattice RP or SAP with  the lattice SAP. In the case of zero radius, i.e. for $r_{\rm ex}=0$, the cylindrical SAP model reduces to the model of equilateral RP. For $r_{\rm ex}=1/8$ it becomes rather close to the lattice SAP.

\section*{Acknowledgements} 
One of the authors (T.D.) would like to thank many participants of the workshop at ICTP 
for helpful discussion: the Workshop on Knots and Links in Biological and Soft Matter Systems, September 26-30, 2016, Trieste, Italy.   The present research is partially supported by the Grant-in-Aid for Scientific Research No. 26310206.

%%%%%%%%%%%%%%%%%%%%%%%%%%%%%%%%%%%%%%%%%%%%%%%%%%%%%%%%%%%%%%%%%%%%%%%%%%%%
\appendix

\section{Best estimates of fitting parameters for some knots} 
 
\subsection{For the trivial knot}

We denote  the knotting probability of the trivial knot for the cylindrical SAP  of $N$ hard cylindrical segments with radius $r_{\rm ex}$ by $P_{0_1}(N, r_{\rm ex})$. 
We also call it the unknotting probability of the SAP. 

It was shown that the unknotting probability decays exponentially with respect to segment number  $N$ for a model of random polygons \cite{Michels-Wiegel} and for the bead-rod model 
of SAP with different bead radii \cite{Koniaris-Muthukumar}. We express the unknotting probability as a function of $N$ most simply by  
\begin{equation}
P_{0_1}(N,r_{\rm ex})=C_0\exp(-{N}/{N_0}) \, . 
\label{eq:unknot}
\end{equation}

There are two fitting parameters in Eq. (\ref{eq:unknot}): $N_0$ and $C_0$, which are 
different from the four parameters employed in the research of Ref. \cite{UD2015}.  

The best estimates of $N_0$ and $C_0$ are listed in Table \ref{tab:knot01}. 
We call $N_{0}$ the characteristic length for the knotting probability of the trivial knot and $C_0$ the coefficient of the knotting probability of the unknot. 

\begin{table}[htbp] 
\begin{center}
\begin{tabular}{c|ccc}
\hline 
$r_{\rm ex}$ & $C_0$ &$N_{0_1}$ & $\chi^2/{\rm DF}$ \\ \hline 
0 & $1.0531 \pm 0.0034$ & $245.89 \pm 0.58$ & $2.36$  \\ 
0.005 & $1.0045 \pm 0.0061$ & $356.0 \pm 1.5$ & $9.01$  \\ 
0.01 & $0.9909 \pm 0.0056$ & $483.4 \pm 2.1$ & $10.64$  \\ 
0.02 & $0.9815 \pm 0.0039$ & $817.7 \pm 3.0$ & $8.38$  \\ 
0.03 & $0.9852 \pm 0.0026$ & $1290.2 \pm 4.0$ & $5.55$  \\ 
0.04 & $0.9865 \pm 0.0017$ & $1965.6 \pm 4.8$ & $3.12$  \\ 
0.05 & $0.9905 \pm 0.0012$ & $2903.1 \pm 6.0$ & $1.90$  \\ 
0.06 & $0.99262 \pm 0.00099$ & $4205. \pm 10.$ & $1.54$  \\ 
0.08 & $0.99648 \pm 0.00079$ & $8290. \pm 23.$ & $0.80$  \\ 
0.1 & $0.99789 \pm 0.00044$ & $15348. \pm 47.$ & $0.43$  \\ 
\hline 
\end{tabular} 
\caption{Best estimates of the parameters in Eq. (\ref{eq:unknot}) for the knotting probability of the trivial knot (the unknot) $0_1$ 
and the $\chi^2$ values per DF. }
\label{tab:knot01}
\end{center}
\end{table}

\subsection{For nontrivial knots}

% -------------------------------------------------
% Figure 9.knotting probabilities of 31#31#31 knot
%
\begin{figure}[htbp]
\begin{center}
  \includegraphics[width=0.80\hsize]{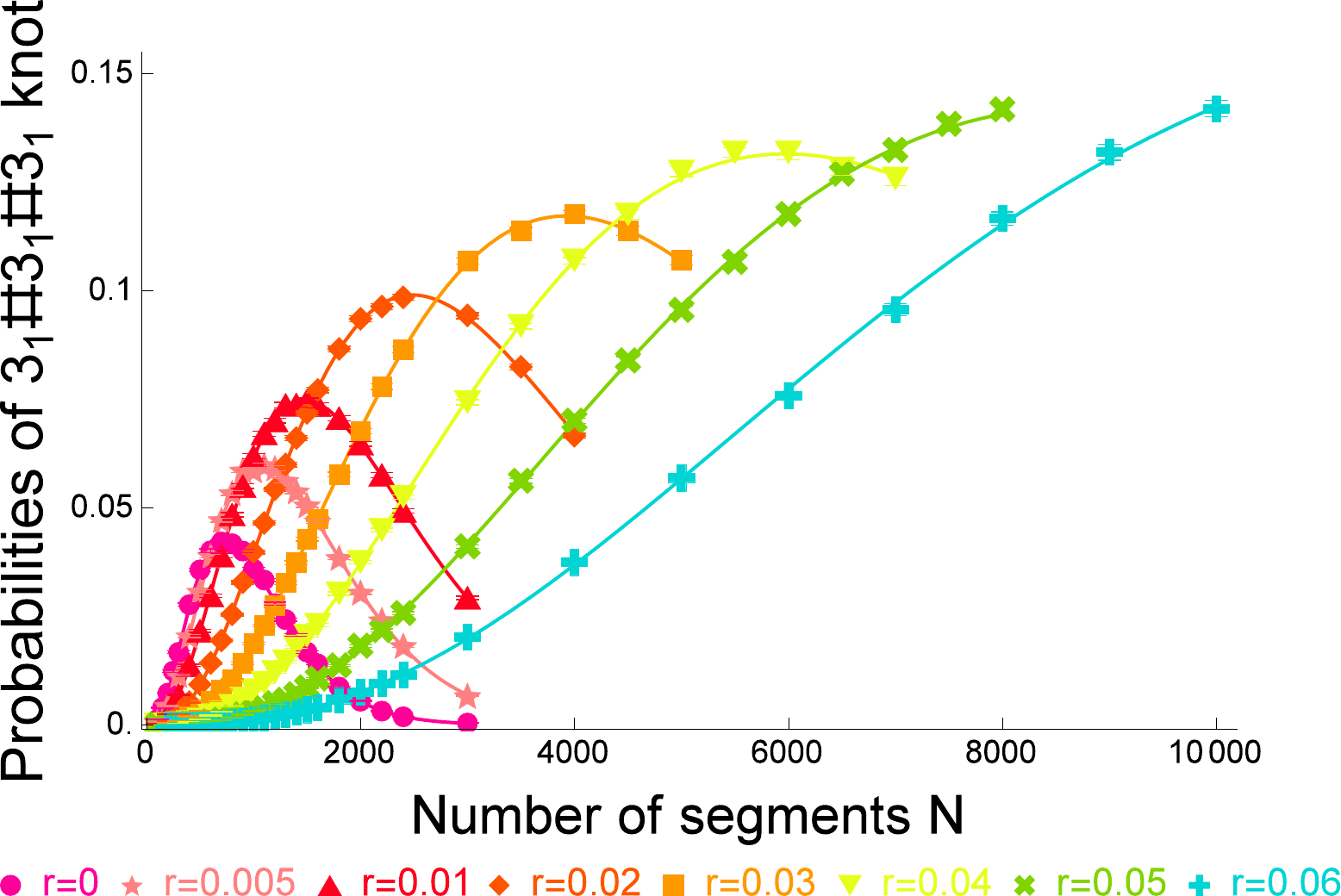}
  \caption{Knotting probability of composite knot $3_1 \sharp 3_1 \sharp 3_1$ versus the number of segments $N$.  }
   \label{fig:313131}
\end{center}
\end{figure}
%-----------------------------------------------------------------------
The knotting probabilities of composite knot $3_1 \sharp 3_1 \sharp 3_1$ 
are plotted against segment number $N$ for different values of radius $r_{\rm ex}$. 
The best estimates of the parameters for Eq. (\ref{eq:4formula}) are listed in Table \ref{tab:313131}. 

\begin{table*}[htbp] 
\center
\begin{tabular}{c|ccccc}
\hline 
$r_{\rm ex}$ & $C_K$ & $m(K)$  & $N_{K}$ & $\Delta N(K)$ & $\chi^2/{\rm DF}$ \\ \hline  
\multicolumn{6}{c}{ Knot $5_1$} \\ \hline 
0 & $0.04376 \pm 0.00038$ & $0.897 \pm 0.059$ & $255.8 \pm 5.5$ & $38.3 \pm 
7.7$ & $1.13$  \\ 
0.005 & $0.03952 \pm 0.00057$ & $0.798 \pm 0.061$ & $386. \pm 11.$ & $45.8 \pm 
9.0$ & $1.71$  \\ 
0.01 & $0.03595 \pm 0.00021$ & $0.944 \pm 0.038$ & $495.4 \pm 8.5$ & $25.4 \pm 
6.6$ & $0.57$  \\ 
0.02 & $0.02899 \pm 0.00026$ & $0.923 \pm 0.043$ & $843. \pm 21.$ & $32.9 \pm 
9.2$ & $1.36$  \\ 
0.03 & $0.02307 \pm 0.00021$ & $0.895 \pm 0.032$ & $1343. \pm 33.$ & $43.2 \pm 
7.7$ & $0.95$  \\ 
0.04 & $0.01881 \pm 0.00024$ & $0.929 \pm 0.041$ & $2006. \pm 74.$ & $41. \pm 
12.$ & $1.58$  \\ 
0.05 & $0.01535 \pm 0.00015$ & $0.932 \pm 0.028$ & $2974. \pm 95.$ & $49.1 \pm 
8.5$ & $0.71$  \\ 
0.06 & $0.01283 \pm 0.00024$ & $0.938 \pm 0.044$ & $4360. \pm 280.$ & $55. \pm 
14.$ & $1.49$  \\ 
0.08 & $0.0095 \pm 0.0003$ & $1.022 \pm 0.067$ & $7800. \pm 1100.$ & $31. \pm 
26.$ & $0.93$  \\ 
0.1 & $0.0074 \pm 0.0011$ & $1.086 \pm 0.095$ & $13000. \pm 4300.$ & $57. \pm 
37.$ & $1.41$  \\ 
%\end{tabular} 
%\caption{Best estimates of the parameters in Eq. (\ref{eq:4formula}) applied to the Knotting %probability of knot $5_1$ for cylindrical SAP with radius $r_{\rm ex}$.} \label{tab:51} \end{table*} 
% 
%5_2 knot \begin{table*}[htbp] \center
%\begin{tabular}{c|ccccc} 
\hline 
%$r_{\rm ex}$ & $C_K$  & $m(K)$  &$N_{K}$  & $\Delta N(K)$  & $\chi^2/{\rm DF}$ \\ \hline 
\multicolumn{6}{c}{ Knot $5_2$} \\ \hline
0 & $0.07693 \pm 0.00041$ & $0.892 \pm 0.034$ & $250.8 \pm 3.2$ & $41.8 \pm 
4.3$ & $0.70$  \\ 
0.005 & $0.06943 \pm 0.00048$ & $0.871 \pm 0.039$ & $371.0 \pm 6.1$ & $38.2 \pm 
5.8$ & $1.04$  \\ 
0.01 & $0.06306 \pm 0.00032$ & $0.959 \pm 0.033$ & $482.1 \pm 7.1$ & $26.5 \pm 
5.6$ & $0.75$  \\ 
0.02 & $0.04985 \pm 0.00029$ & $0.906 \pm 0.027$ & $844. \pm 14.$ & $35.0 \pm 
5.6$ & $0.90$  \\ 
0.03 & $0.03959 \pm 0.00032$ & $0.92 \pm 0.031$ & $1339. \pm 32.$ & $41.1 \pm 
7.5$ & $1.45$  \\ 
0.04 & $0.03200 \pm 0.00019$ & $0.951 \pm 0.021$ & $1971. \pm 37.$ & $35.5 \pm 
6.3$ & $0.66$  \\ 
0.05 & $0.02578 \pm 0.00020$ & $0.934 \pm 0.023$ & $3023. \pm 80.$ & $46.7 \pm 
7.4$ & $0.79$  \\ 
0.06 & $0.02119 \pm 0.00029$ & $0.938 \pm 0.033$ & $4340. \pm 210.$ & $55. \pm 
11.$ & $1.36$  \\ 
0.08 & $0.01431 \pm 0.00043$ & $0.942 \pm 0.054$ & $8100. \pm 1000.$ & $63. \pm 
16.$ & $1.51$  \\ 
0.1 & $0.0108 \pm 0.0010$ & $1.022 \pm 0.064$ & $13300. \pm 3100.$ & $67. \pm 
23.$ & $1.11$  \\ 
\hline 
%\end{tabular} 
%\caption{Best estimates of the parameters in Eq. (\ref{eq:4formula}) applied to the knotting probability of knot $5_2$ for cylindrical SAP with radius $r_{\rm ex}$.} \label{tab:52} \end{table*} 
%
%6_1 knot 
% \begin{table*}[htbp] 
%\center \begin{tabular}{c|ccccc} 
%\hline
%$r_{\rm ex}$ & $C_K$  & $m(K)$  &$N_{K}$  & $\Delta N(K)$ & $\chi^2/{\rm DF}$ \\ \hline 
\hline 
\multicolumn{6}{c}{ Knot $6_1$} \\ \hline
0 & $0.02255 \pm 0.00026$ & $0.808 \pm 0.045$ & $271.4 \pm 5.3$ & $60.4 \pm 
5.$ & $0.72$  \\ 
0.005 & $0.01888 \pm 0.00026$ & $0.872 \pm 0.063$ & $383. \pm 11.$ & $57.2 \pm 
7.5$ & $1.14$  \\ 
0.01 & $0.01596 \pm 0.00011$ & $0.921 \pm 0.034$ & $501.8 \pm 8.7$ & $50.4 \pm 
4.7$ & $0.32$  \\ 
0.02 & $0.01092 \pm 0.00016$ & $0.891 \pm 0.056$ & $875. \pm 33.$ & $52. \pm 
10.$ & $1.21$  \\ 
0.03 & $0.007856 \pm 0.000080$ & $1.005 \pm 0.049$ & $1243. \pm 43.$ & $45. \pm 
11.$ & $0.70$  \\ 
0.04 & $0.00544 \pm 0.00012$ & $0.892 \pm 0.055$ & $2050. \pm 120.$ & $69. \pm 
12.$ & $1.26$  \\ 
0.05 & $0.004112 \pm 0.000070$ & $0.987 \pm 0.058$ & $2840. \pm 170.$ & $44. \pm 
18.$ & $0.73$  \\ 
0.06 & $0.003126 \pm 0.000086$ & $1.035 \pm 0.082$ & $3960. \pm 410.$ & $41. 
\pm 29.$ & $1.01$  \\ 
%\end{tabular}  \end{table} 
 %6_2 knot 
%\begin{table}[htbp] 
%\begin{tabular}{c|ccccc} 
\hline 
\multicolumn{6}{c}{ Knot $6_2$} \\ \hline 
%$r_{\rm ex}$ & $C_K$  & $m(K)$  &$N_{K}$  & $\Delta N(K)$  & $\chi^2/{\rm DF}$ \\ \hline 
0 & $0.02420 \pm 0.00028$ & $0.963 \pm 0.082$ & $249.9 \pm 7.2$ & $38. \pm 10.$ & 
$1.20$  \\ 
0.005 & $0.02017 \pm 0.00023$ & $0.953 \pm 0.068$ & $360. \pm 10.$ & $44.2 \pm 
8.9$ & $1.01$  \\ 
0.01 & $0.01675 \pm 0.00023$ & $0.944 \pm 0.073$ & $492. \pm 18.$ & $48. \pm 
10.$ & $1.46$  \\ 
0.02 & $0.01141 \pm 0.00015$ & $0.862 \pm 0.043$ & $859. \pm 26.$ & $62.9 \pm 
6.9$ & $0.93$  \\ 
0.03 & $0.00782 \pm 0.00016$ & $0.873 \pm 0.058$ & $1373. \pm 70.$ & $68. \pm 
10.$ & $1.61$  \\ 
0.04 & $0.005412 \pm 0.000084$ & $1.036 \pm 0.069$ & $1785. \pm 99.$ & $43. \pm 
18.$ & $1.17$  \\ 
0.05 & $0.003926 \pm 0.000086$ & $1.012 \pm 0.077$ & $2710. \pm 220.$ & $52. 
\pm 22.$ & $1.30$  \\ 
0.06 & $0.003132 \pm 0.000068$ & $1.222 \pm 0.097$ & $3430. \pm 310.$ & $-16. 
\pm 41.$ & $0.68$  \\ 
%\end{tabular} 
%\end{table}  
%6_3 knot 
%\begin{table}[htbp] 
%\begin{tabular}{c|ccccc} 
\hline 
\multicolumn{6}{c}{ Knot $6_3$} \\ \hline 
%$r_{\rm ex}$ & $C_K$  & $m(K)$  &$N_{K}$  & $\Delta N(K)$  & $\chi^2/{\rm DF}$ \\ \hline 
0 & $0.01593 \pm 0.00024$ & $0.832 \pm 0.065$ & $260.5 \pm 7.2$ & $60.3 \pm 
6.9$ & $0.92$  \\ 
0.005 & $0.01319 \pm 0.00020$ & $1.063 \pm 0.084$ & $337. \pm 11.$ & $35. \pm 
11.$ & $0.79$  \\ 
0.01 & $0.01057 \pm 0.00017$ & $0.943 \pm 0.086$ & $489. \pm 21.$ & $49. \pm 
12.$ & $1.30$  \\ 
0.02 & $0.00666 \pm 0.00014$ & $0.863 \pm 0.068$ & $869. \pm 42.$ & $64. \pm 
11.$ & $1.35$  \\ 
0.03 & $0.00430 \pm 0.00011$ & $0.885 \pm 0.084$ & $1378. \pm 92.$ & $43. \pm 
20.$ & $1.24$  \\ 
0.04 & $0.003081 \pm 0.000080$ & $0.929 \pm 0.074$ & $1960. \pm 140.$ & $68. \pm 
16.$ & $1.21$  \\ 
0.05 & $0.002045 \pm 0.000081$ & $0.829 \pm 0.074$ & $3220. \pm 340.$ & $84. 
\pm 13.$ & $1.34$  \\ 
0.06 & $0.001581 \pm 0.000055$ & $1.22 \pm 0.15$ & $3070. \pm 420.$ & $19. \pm 
55.$ & $1.07$  \\ 
\hline 
\end{tabular} 
\caption{
Best estimates of the parameters in Eq. (\ref{eq:4formula}) applied to the knotting probabilities of knots $5_1$, $5_2$, $6_1$, $6_2$, and $6_3$ for the cylindrical SAP with radius $r_{\rm ex}$.  }
\label{tab:56}
\end{table*} 

%7_1 knot 
\begin{table*}[htbp] \center
\begin{tabular}{c|ccccc} \hline
$r_{\rm ex}$ & $C_K$  & $m(K)$  & $N_{K}$  & $\Delta N(K)$  & $\chi^2/{\rm DF}$ \\ \hline 
\multicolumn{6}{c}{ Knot $7_1$ } \\ \hline 
0 & $0.003015 \pm 0.000072$ & $0.97 \pm 0.16$ & $256. \pm 15.$ & $43. \pm 19.$ & 
$0.65$  \\ 
0.005 & $0.00241 \pm 0.000076$ & $0.84 \pm 0.12$ & $371. \pm 22.$ & $71. \pm 
11.$ & $0.69$  \\ 
0.01 & $0.00182 \pm 0.00016$ & $1.29 \pm 0.30$ & $431. \pm 47.$ & $12. \pm 44.$ & 
$1.32$  \\ 
0.02 & $0.001158 \pm 0.000054$ & $1.16 \pm 0.25$ & $744. \pm 95.$ & $24. \pm 
49.$ & $1.33$  \\ 
0.03 & $0.000757 \pm 0.000027$ & $1.02 \pm 0.16$ & $1320. \pm 170.$ & $55. \pm 
34.$ & $0.88$  \\ 
0.04 & $0.000429 \pm 0.000039$ & $0.84 \pm 0.21$ & $2410. \pm 600.$ & $68. \pm 
50.$ & $1.36$  \\ 
\hline 
%\end{tabular} \caption{$7_1$ knot } \end{table} 
% %7_2 knot 
% \begin{table}[htbp] \begin{tabular}{c|ccccc} 
\multicolumn{6}{c}{ Knot $7_2$} \\ \hline 
%$r_{\rm ex}$ & $C_K$  & $m(K)$  & $N_{K}$ &  $\Delta N(K)$ & $\chi^2/{\rm DF}$ \\ \hline 
0 & $0.00677 \pm 0.00019$ & $1.1 \pm 0.15$ & $238. \pm 13.$ & $45. \pm 16.$ & 
$1.27$  \\ 
0.005 & $0.00534 \pm 0.00022$ & $1.09 \pm 0.20$ & $339. \pm 26.$ & $43. \pm 24.$ & 
$2.18$  \\ 
0.01 & $0.004117 \pm 0.000069$ & $1.037 \pm 0.091$ & $478. \pm 21.$ & $53. \pm 
11.$ & $0.61$  \\ 
0.02 & $0.002458 \pm 0.000082$ & $0.89 \pm 0.12$ & $878. \pm 72.$ & $57. \pm 
21.$ & $1.38$  \\ 
0.03 & $0.001609 \pm 0.000052$ & $0.94 \pm 0.11$ & $1360. \pm 120.$ & $72. \pm 
18.$ & $1.12$  \\ 
0.04 & $0.001018 \pm 0.000035$ & $1.16 \pm 0.18$ & $1710. \pm 230.$ & $45. \pm 
43.$ & $1.43$  \\ 
\hline 
%\end{tabular} \caption{$7_2$ knot }
%\end{table} 
 %7_3 knot    
%\begin{table}[htbp] 
%\begin{tabular}{c|ccccc} 
\multicolumn{6}{c}{ Knot $7_3$} \\ \hline 
%$r_{\rm ex}$ & $C_K$  & $m(K)$  & $N_{K}$  & $\Delta N(K)$  & $\chi^2/{\rm DF}$ \\ \hline 
0 & $0.00550 \pm 0.00022$ & $1.16 \pm 0.18$ & $228. \pm 14.$ & $39. \pm 18.$ & 
$1.17$  \\ 
0.005 & $0.00413 \pm 0.00010$ & $0.95 \pm 0.13$ & $366. \pm 22.$ & $55. \pm 15.$ & 
$0.99$  \\ 
0.01 & $0.003430 \pm 0.000077$ & $1.04 \pm 0.13$ & $472. \pm 27.$ & $40. \pm 
18.$ & $0.81$  \\ 
0.02 & $0.001995 \pm 0.000059$ & $0.845 \pm 0.088$ & $877. \pm 57.$ & $67. \pm 
14.$ & $0.75$  \\ 
0.03 & $0.001255 \pm 0.000059$ & $0.84 \pm 0.11$ & $1420. \pm 150.$ & $83. \pm 
13.$ & $1.28$  \\ 
0.04 & $0.00081 \pm 0.000033$ & $0.927 \pm 0.094$ & $1920. \pm 200.$ & $92.6 
\pm 9.4$ & $0.89$  \\  
%\end{tabular} 
%\caption{Best estimates of the parameters in Eq. (\ref{eq:4formula}) applied to the knotting probabilities of knots $7_1$, $7_2$ and $7_3$ for cylindrical SAP with radius $r_{\rm ex}$. }
%\label{tab:7-123} \end{table*}  
%7_4 knot 
% \begin{table*}[htbp] \center
%\begin{tabular}{c|ccccc}
\hline 
%$r_{\rm ex}$ & $C_K$  & $m(K)$  &$N_{K}$  & $\Delta N(K)$  & $\chi^2/{\rm DF}$ \\ \hline 
\multicolumn{6}{c}{ Knot $7_4$ } \\ \hline  
0 & $0.00278 \pm 0.00018$ & $1.23 \pm 0.25$ & $239. \pm 18.$ & $19. \pm 29.$ & 
$0.8$  \\ 
0.005 & $0.002228 \pm 0.000062$ & $0.99 \pm 0.18$ & $373. \pm 28.$ & $42. \pm 
23.$ & $0.68$  \\ 
0.01 & $0.001783 \pm 0.000050$ & $0.95 \pm 0.16$ & $504. \pm 40.$ & $38. \pm 
26.$ & $0.68$  \\ 
0.02 & $0.001011 \pm 0.000030$ & $0.93 \pm 0.10$ & $816. \pm 58.$ & $76. \pm 12.$ & 
$0.55$  \\ 
0.03 & $0.000659 \pm 0.000027$ & $1.05 \pm 0.19$ & $1180. \pm 160.$ & $62. \pm 
32.$ & $1.09$  \\ 
0.04 & $0.000378 \pm 0.000041$ & $1.38 \pm 0.42$ & $1670. \pm 370.$ & $-90. \pm 
150.$ & $0.88$  \\ 
\hline 
%\end{tabular}  \caption{$7_4$ knot } \end{table} 
 %7_5 knot 
% \begin{table}[htbp] \begin{tabular}{c|ccccc} 
\multicolumn{6}{c}{ Knot $7_5$ } \\ \hline 
%$r_{\rm ex}$ & $C_K$  & $m(K)$  &$N_{K}$  & $\Delta N(K)$  & $\chi^2/{\rm DF}$ \\ \hline 
0 & $0.00772 \pm 0.00016$ & $1.06 \pm 0.12$ & $246. \pm 10.$ & $44. \pm 13.$ & 
$0.88$  \\ 
0.005 & $0.00625 \pm 0.00014$ & $0.98 \pm 0.12$ & $353. \pm 19.$ & $52. \pm 
14.$ & $1.15$  \\ 
0.01 & $0.004647 \pm 0.000075$ & $0.906 \pm 0.071$ & $506. \pm 19.$ & $58.9 \pm 
9.$ & $0.49$  \\ 
0.02 & $0.002754 \pm 0.000045$ & $1.053 \pm 0.09$ & $775. \pm 40.$ & $44. \pm 
16.$ & $0.61$  \\ 
0.03 & $0.001645 \pm 0.000037$ & $1.12 \pm 0.12$ & $1104. \pm 86.$ & $53. \pm 
23.$ & $0.9$  \\ 
0.04 & $0.001065 \pm 0.000036$ & $1.23 \pm 0.16$ & $1580. \pm 170.$ & $67. \pm 
37.$ & $1.28$  \\ 
\hline 
%\end{tabular} 
%\caption{Best estimates of the parameters in Eq. (\ref{eq:4formula}) applied to the knotting probabilities of knots $7_4$ and $7_5$ for cylindrical SAP with radius $r_{\rm ex}$. }
%\label{tab:7-45} \end{table*}  
%\begin{table*}[htbp] \center
%\begin{tabular}{c|ccccc} \hline 
%$r_{\rm ex}$ & $C_K$ & $m(K)$ & $N_{K}$  & $\Delta N(K)$  & $\chi^2/{\rm DF}$ \\ \hline 
\multicolumn{6}{c}{ Knot $7_6$} \\ 
\hline 
0 & $0.00919 \pm 0.00018$ & $1.04 \pm 0.12$ & $248. \pm 10.$ & $46. \pm 12.$ & 
$1.14$  \\ 
0.005 & $0.00705 \pm 0.00011$ & $0.958 \pm 0.081$ & $366. \pm 13.$ & $54.5 \pm 
9.3$ & $0.63$  \\ 
0.01 & $0.00549 \pm 0.00010$ & $1.08 \pm 0.10$ & $468. \pm 22.$ & $44. \pm 14.$ & 
$0.82$  \\ 
0.02 & $0.003167 \pm 0.000077$ & $1.12 \pm 0.14$ & $738. \pm 51.$ & $22. \pm 
28.$ & $1.12$  \\ 
0.03 & $0.001907 \pm 0.000049$ & $1.036 \pm 0.096$ & $1241. \pm 97.$ & $80. \pm 
14.$ & $1.19$  \\ 
0.04 & $0.001165 \pm 0.000032$ & $1.1 \pm 0.14$ & $1780. \pm 200.$ & $39. \pm 
37.$ & $0.98$  \\ 
\hline 
%\end{tabular}  \caption{$7_6$ knot }\end{table} 
 %7_7 knot 
% \begin{table}[htbp] \begin{tabular}{c|ccccc} 
\multicolumn{6}{c}{ Knot $7_7$} \\ \hline 
%$r_{\rm ex}$ & $C_K$  & $m(K)$  &$N_{K}$  & $\Delta N(K)$  & $\chi^2/{\rm DF}$ \\ \hline 
0 & $0.00602 \pm 0.00050$ & $1.42 \pm 0.21$ & $246. \pm 15.$ & $22. \pm 23.$ & 
$1.77$  \\ 
0.005 & $0.00475 \pm 0.00016$ & $1.26 \pm 0.12$ & $388. \pm 18.$ & $29. \pm 
16.$ & $0.7$  \\ 
0.01 & $0.003500 \pm 0.000084$ & $1.19 \pm 0.11$ & $526. \pm 27.$ & $32. \pm 17.$ & 
$0.58$  \\ 
0.02 & $0.002139 \pm 0.000046$ & $1.12 \pm 0.11$ & $847. \pm 59.$ & $61. \pm 
17.$ & $0.93$  \\ 
0.03 & $0.001099 \pm 0.000048$ & $0.97 \pm 0.16$ & $1390. \pm 190.$ & $66. \pm 
28.$ & $1.43$  \\ 
0.04 & $0.000673 \pm 0.000034$ & $1.28 \pm 0.24$ & $1600. \pm 260.$ & $41. \pm 
61.$ & $1.3$  \\ 
\hline 
\end{tabular} 
\caption{
Best estimates of the parameters in Eq. (\ref{eq:4formula}) applied to the knotting probabilities of knots $7_1$, $7_2$, $7_3$, $7_4$, $7_5$,  $7_6$ and $7_7$ for the cylindrical SAP with radius $r_{\rm ex}$. }
\label{tab:7} 
\end{table*}

%%%%%%%%%%%%%%%%%%%%%%%%%%% 

% ---------------------------------------------------------
% Table 6
\begin{table*}[htbp] 
\begin{center}
\begin{tabular}{c|ccccc}
\hline  
$r_{\rm ex}$ & $C_K$ & $m(K)$  & $N_K$ & $\Delta N(K)$ & $\chi^2$/DF \\
\hline 
0 & $0.0390 \pm 0.0023$ & $2.806 \pm 0.058$ & $257.8 \pm 2.9$ & $14.1 \pm 5.2$ & 
$1.69$  \\ 
0.005 & $0.0546 \pm 0.0024$ & $2.810 \pm 0.044$ & $379.7 \pm 3.9$ & $15.4 \pm 
4.9$ & $1.11$  \\ 
0.01 & $0.0651 \pm 0.0023$ & $2.853 \pm 0.035$ & $511.0 \pm 4.8$ & $16.6 \pm 
4.6$ & $0.73$  \\ 
0.02 & $0.0848 \pm 0.0026$ & $2.869 \pm 0.03$ & $859.8 \pm 8.2$ & $19.2 \pm 
5.6$ & $0.71$  \\ 
0.03 & $0.0944 \pm 0.0044$ & $2.927 \pm 0.044$ & $1336. \pm 23.$ & $15. \pm 
10.$ & $1.17$  \\ 
0.04 & $0.1085 \pm 0.0042$ & $2.904 \pm 0.035$ & $2049. \pm 32.$ & $25. \pm 
11.$ & $0.81$  \\ 
0.05 & $0.1106 \pm 0.0075$ & $2.960 \pm 0.055$ & $2948. \pm 84.$ & $20. \pm 21.$ & 
$1.16$  \\ 
0.06 & $0.139 \pm 0.011$ & $2.851 \pm 0.057$ & $4660. \pm 180.$ & $61. \pm 25.$ & 
$0.87$  \\ 
\hline 
\end{tabular} 
\end{center} 
\caption{Best estimates of the parameters of Eq. (\ref{eq:4formula}) for composite knot $3_1\#3_1\#3_1$. }  
\label{tab:313131}
\end{table*}

%***********************************************************************
% References
%***********************************************************************


\newpage


\begin{thebibliography}{99}

\bibitem{Kramers}  H.A. Kramers, J. Chem. Phys. {\bf 14}, 415 (1946). 

\bibitem{Semlyen} 
{\it Cyclic Polymers}, ed. J. A. Semlyen, (Elsevier Applied Science Publishers, New York, 1986); 2nd Ed. (Kluwer Academic Publ., Dordrecht, 2000)     

%[1] Cyclic Polymers, ed. J. A. Semlyen, (Elsevier Applied Science Publishers, New York, 1986); 2nd Ed. (Kluwee Academic Publ., Dordrecht, 2000)     


\bibitem{Bates} A.D.~Bates and A.~Maxwell, {\it DNA Topology} 
(Oxford Univ. Press, 2005). 

%[3] A.D..Bates and A. Maxwell, DNA Topology (Oxford Univ. Press, 2005, Oxford).


\bibitem{Vinograd} J. Vinograd, J. Lebowitz, R. Radloff, R. Watson, and P. Laipis, Proc. Natl. Acad. Sci. (U.S.) {\bf 53}, 1104 (1965).  

%[2]  J. Vinograd, J. Lebowitz, R. Radloff, R. Watson, and P. Laipis, Proc. Natl. Acad. Sci. (U.S.) 53, 1104 (1965).  


\bibitem{Nature-trefoil}
M.A. Krasnow, A. Stasiak, S.J. Spengler, F. Dean, T. Koller and N.R. Cozzarelli, %Determination of the absolute handedness of knots and catenanes of DNA, 
{Nature}, {\bf 304},  559 (1983). 
%559--560. 

\bibitem{DNAknots} F.B. Dean, A. Stasiak,  T. Koller and N.R. Cozzarelli, 
%Duplex DNA Knots Produced by {\it Escherichia coli} Topoisomerase I,  
{ J. Biol. Chem.},  {\bf 260}, 4975 (1985). 
%--4983. 





\bibitem{Taylor} W.R. Taylor, 
%A deeply knotted protein structure and how it might fold, 
{Nature}, {\bf 406}, 916 (2000).  
%-- 919 
 
%[4] W.R. Taylor, A deeply knotted protein structure and how it might fold, Nature, 406, 916 (2000).  

\bibitem{Craik} D. J. Craik, Science {\bf 311}, 1563 (2006). 
%[5] D. J. Craik, Science 311, 1563 (2006).   


\bibitem{Woltering} J.J. Danon et al., Science {\bf 355}, 159-162 (2017).   

\bibitem{Tezuka2000} H. Oike, H. Imaizumi, T. Mouri, Y. Yoshioka, A. Uchibori, and Y. Tezuka, 
J. Am. Chem. Soc. 122, 9592 (2000).  

%[6] H. Oike, H. Imaizumi, T. Mouri, Y. Yoshioka, A. Uchibori, and Y. Tezuka, J. Am. Chem. Soc. 122, 9592 (2000).  



\bibitem{Tezuka2001} Y. Tezuka and H. Oike, J. Am. Chem. Soc. {\bf 123}, 11570 (2001).

%[7] Y. Tezuka and H. Oike, J. Am. Chem. Soc. 123, 11570 (2001).


\bibitem{Grubbs} C. W. Bielawski, D. Benitez and R. H. Grubbs, Science \textbf{297}, 2041--2044 (2002).

%[8] C. W. Bielawski, D. Benitez and R. H. Grubbs, Science 297, 2041 (2002).


\bibitem{Takano05}
D. Cho, K. Masuoka, K. Koguchi, T. Asari, D. Kawaguchi, A. Takano 
and Y. Matsushita, Polymer Journal \textbf{37}, 506--511 (2005).

%[9] D. Cho, K. Masuoka, K. Koguchi, T. Asari, D. Kawaguchi, A. Takano and Y. Matsushita, Polymer Journal 37, 506 (2005).



\bibitem{Takano07}
A. Takano, Y. Kushida, K. Aoki, K. Masuoka, K. Hayashida, 
D. Cho, D. Kawaguchi and Y. Matsushita, 
Macromolecules \textbf{40}, 679--681 (2007).

%[10] A. Takano, Y. Kushida, K. Aoki, K. Masuoka, K. Hayashida, D. Cho, D. Kawaguchi and Y. Matsushita, Macromolecules 40, 679--681 (2007).


\bibitem{Grayson} B. A. Laurent and S. Grayson, 
J. Am. Chem. Soc. {\bf 128}, 4238--4239 (2006). 

%[11]  B. A. Laurent and S. Grayson, J. Am. Chem. Soc. 128, 4238 (2006).  



\bibitem{Tezuka2010}  N. Sugai, H. Heguri, K. Ohta, Q. Meng, 
T. Yamamoto and Y. Tezuka,  J. Am. Chem. Soc. {\bf 132}, 14790--14802 (2010) 

%[12] N. Sugai, H. Heguri, K. Ohta, Q. Meng, T. Yamamoto and Y. Tezuka,  J. Am. Chem. Soc. 132, 14790 (2010) 

\bibitem{Tezuka2011} N. Sugai, H. Heguri, T. Yamamoto and Y. Tezuka, 
J. Am. Chem. Soc. {\bf 133},  19694--19697 (2011).  
 
\bibitem{Tezuka-book} {\it Topological Polymer Chemistry: 
Progress in cyclic polymers in syntheses, properties and functions}, 
ed. by Y. Tezuka, (World Scientific Publ., Singapore, 2013). 





%%%%%%%%%%%%%%%%%%%%%%%%%%%%%%%%%%%%%%%%%%%%%%%%%%%%%%%
% trivial knot probability 
%

\bibitem{Vologodskii} A.~V. Vologodskii, A.V. Lukashin, 
M.~D. Frank-Kamenetskii, and V.~V. Anshelevich,
%The knot probability in statistical mechanics of polymer chains,  
{ Sov. Phys. JETP}, {\bf 39}, 1059 (1974).

\bibitem{Michels-Wiegel} J.~P.~J. Michels and F.~W. Wiegel, Phys. Lett. A {\bf 90}, 381 (1982).  
%Proc. R. Soc. London, Ser. A, \textbf{403}, 269 (1996).

\bibitem{Janse van Rensburg} E.~J. Janse van Rensburg and S.~G. Whittington, 
J. Phys. A: Math. Gen. \textbf{23}, 3573 (1990)

\bibitem{Koniaris-Muthukumar} K. Koniaris and M. Muthukumar, 
%Knottedness in Ring Polymers, 
Phys. Rev. Lett. \textbf{66}, 2211 (1991).



\bibitem{knotP} T.~Deguchi and K.~Tsurusaki, 
%Topology of Closed Random Polygons, 
{ J. Phys. Soc. Jpn.}, {\bf 62}, 1411 (1993).  
%1411-1414. 


\bibitem{JKTR} T.~Deguchi and K.~Tsurusaki,  
%A Statistical Study of Random Knotting Using the Vassiliev Invariants, 
{ J. Knot Theory Ramif.}, {\bf 3}, 321 (1994).
%321-353. 



\bibitem{TD95} K. Tsurusaki and T. Deguchi,  
%Fractions of Particular knots in Gaussian Random Polygons, \\
J. Phys. Soc. Jpn. {\bf 64}, 1506 (1995). 
%1506-1518.


%%%%%%%%%%%%%%%%%%%%%%%%%%%%%%%%%%%%%%%%%
%
% Deguchi-Tsurusaki1997,Orlandini1996,Orlandini1998,PLA2000,Yao,Marcone,Stella,Rechnitzer,TubianaUD2015

\bibitem{Deguchi-Tsurusaki1997} T. Deguchi and K. Tsurusaki,  
%Universality of random knotting, 
Phys. Rev. E. \textbf{55},  6245 (1997).



\bibitem{Orlandini1996} E. Orlandini, M.~C. Tesi, E.~J. Janse van Rensburg, 
and S.G. Whittington, 
%Entropic exponents of lattice polygons with specified knot type, 
{ J. Phys. A: Math. Gen.}, {\bf 29}, L299 (1996). 
%L299--L303. 

\bibitem{Orlandini1998} E. Orlandini, M.~C. Tesi, E.~J. Janse van Rensburg, 
and S.~G. Whittington, 
%Asymptotics of knotted lattice polygons, 
{ J. Phys. A: Math. Gen.}, {\bf 31}, 5953 (1998).  
%5953--5967. 



\bibitem{PLA2000} M.~K. Shimamura and T. Deguchi, 
%Characteristic length of random knotting for cylindrical self-avoiding polygons, 
%
 Phys. Lett. A {\bf 274}, 184 (2000).  
%pp. 184-191.  

\bibitem{Katritch00} V. Katritch, W. K. Olsen, A. Vologodskii, J. Dubochet, and A. Stasiak, Phys. Rev. E {\bf 61}, 5545 (2000).  


\bibitem{Yao} A. Yao, H. Matsuda, H. Tsukahara, M.~K. Shimamura, and T. Deguchi, 
%On the dominance of trivial knots among SAP on a cubic lattice, 
J. Phys. A: Math. Gen., {\bf 34}, 7563 (2001).   
%pp. 7563-7577
% (cond-mat/0103365) 


\bibitem{Marcone} B. Marcone, E. Orlandini,  A.~L. Stella and F. Zonta, 
%What is the length of a knot in a polymer
J. Phys. A: Math. Gen., {\bf 38}, L15-L21 (2005).   


\bibitem{Stella} M. Baiesi, E. Orlandini and A.~L. Stella, 
%The entropic cost to tie a knot, 
J. Stat. Mech., P06012 (2010). 

\bibitem{Rechnitzer} E.~J.~Janse van Rensburg and A. Rechnitzer, 
%On the universality of knot probability ratios,  E. 
J. Phys. A: Math. Theor. {\bf 44}, 162002 (2011)  
%(8 page). 



\bibitem{Tubiana} L. Tubiana,  
% 
Phys. Rev. E. \textbf{89},  052602 (2014).

\bibitem{UD2015} E. Uehara and T. Deguchi, J. Phys. Condens. Matter \textbf{27}, 354104 (2015). 








%%%%%%%%%%%%%%%%%%%%%%%%%%%%%%%%%%%%%%
%
%Orlandini1998,Katritch00,Marcone05
%

%\bibitem{Orlandini1998} E. Orlandini, M.~C. Tesi, E.~J. Janse van Rensburg, and S.~G. Whittington, 
%Asymptotics of knotted lattice polygons, { J. Phys. A: Math. Gen.}, {\bf 31}, 5953 (1998).  
%5953--5967. 



%\bibitem{Marcone05} B. Marcone, E. Orlandini,  A.~L. Stella and F. Zonta, J. Phys. A:  {\bf 38}, L15 (2005).   





%%%%%%%%%%%%%%%%%%%%%%%%%%%%%%%%%%%%%%%%%%%%%%%%%%%%
%
%  Random knotting in SAP 
%
%%%%%%%%%%%%%%%%%%%%%%%%%%%%%%%%%%%%%%%%%%%%%%%%%%%%%


\bibitem{Sumners-Whittington} D.~W. Sumners and S. Whittington
{ J. Phys. A: Math. Gen.}, {\bf 21}, 1689 (1988). 
%1689-1694. 


\bibitem{Pippenger} N. Pippenger, 
{Discrete Appl. Math.}, {\bf 25}, 273 (1989). 
%273-278.  



%%%%%%%%%%%%%%%%%%%%%%%%%%%%%%%%%%%%%%%%%%%%%%%%%
%
% Measurement of random knotting probabilities 
% Rybenkov}Shaw-WangPlesa
%

\bibitem{Rybenkov} V.~V. Rybenkov, N.~R. Cozzarelli and A.~V. Vologodskii, 
% Probability of DNA knotting and the effective diameter of the DNA double helix, 
Proc. Natl. Acad. Sci. USA {\bf 90}, 5307 (1993).  


\bibitem{Shaw-Wang} S.~Y. Shaw and J.~C. Wang, 
%Knotting of a DNA Chain During Ring Closure, 
Science {\bf 260}, 533 (1993).  

\bibitem{Plesa} C. Plesa et al., Nature Nanotech. (2016) DOI: 10.1038/NNANO.2016.153 .









%%%%%%%%%%%%%%%%%%%%%%%%%%%%%%%%%%%%%%%%%%%
%
%  Electric double layer theory 
%  

\bibitem{Schellman} J.~A. Schellman and D. Stigter, {Biopolymers} {\bf 16}, 1415 (1977);  
\bibitem{Stigter}  D. Stigter, { Biopolymers} {\bf 16}, 1435 (1977).    



\bibitem{LeBret} M. Le Bret and B.~H. Zimm, { Biopolymers} {\bf 23}, 287 (1984). 



%%%%%%%%%%%%%%%%%%%%%%%%%%%%%%%%%%%%%
%
%  textbooks

\bibitem{Grosberg-book} A. Yu Grosberg and A. R. Khokhlov, {\it Statistical Physics of Macromolecules} (AIP press, 1994, New York).  


\bibitem{Murasugi} K. Murasugi and B. Kurpita, 
{\it Knot Theory and Its Applications} (Birkh{\"a}user, Boston, 1996).   

%%%%%%%%%%%%%%%%%%%%%%%%%%%%%%%%%%%%%%%%%%%%


\bibitem{Grosberg} A. Grosberg, Macromolecules {\bf 41},  4524 (2008).    
%4524-4527

\bibitem{Jason} J. Cantarella, T. Deguchi and C. Shonkwiler, 
%Probability Theory of Random Polygons from the Quaternionic Viewpoint, 
Comm. Pure Appl. Math. {\bf 67}, 1658-1699 (2014).  
%arXiv:1206.3161, 
%Communications on Pure and Applied Mathematics (2013).   
%Comm. Pure Appl. Math.. doi: 10.1002/cpa.21480 . 





\bibitem{Millett1994} K.C. Millett, 
%Knotting of regular polygons,  
J. Knot Theory Ramif. {\bf 3}, 263 (1994). 
%263-278   

\bibitem{Kapovich1996} M. Kapovich and J.~J. Milson, 
% The symplectic geometry of polygons in Euclidean space, 
J. Differ. Geom. {\bf 44}, 479 (1996).   
%479-513  

\bibitem{Millett2011} S. Alvarado, J.~A. Calvo, K.~C. Millett, 
%The Generation of Random Equilateral Polygons, 
J. Stat. Phys. {\bf 143}, 102 (2011).  
%102-138

%%%%%%%%%%%%%%%%%%%%%%%%%%%%%%%%%%%%%%%%%%%%%%%%%%%%%



\bibitem{Deguchi-Tsurusaki-PLA} T. Deguchi and K. Tsurusaki,  
% A New Algorithm for Numerical Calculation of Link Invariants, 
Phys. Lett. A {\bf 174} (1993) 29-37.

\bibitem{Polyak-Viro} M. Polyak and O. Viro, { IMRN.}, \textbf{No. 11}, 445 (1994).



%%%%%%%%%%%%%%%%%%%%%%%%%%%%%%%%
\bibitem{Rosa11} A. Rosa,  E. Orlandini, L. Tubiana and C. Micheletti, 
% Structure and Dynamics  of Ring Polymers: Entanglement Effects Because of Solution Density and Ring Topology 
Macromolecules {\bf 44},  8668 (2011).   


\end{thebibliography}
\end{document}